\documentclass[a4paper,12pt,american]{article}
\usepackage[utf8]{inputenc}
\usepackage[T1]{fontenc}
\usepackage[american]{babel}
\usepackage{csquotes}
\usepackage[hidelinks]{hyperref}
\usepackage{amsmath,  amsfonts, amssymb}
\usepackage[detect-all]{siunitx}
\usepackage{graphicx}
\usepackage{xcolor}
\usepackage[super]{nth}
\usepackage{bm}
\usepackage{tikz}
\usetikzlibrary{calc}

\usepackage{algotitle}

\ATtitle{Multivariate extensions\\ of the Multilevel Best Linear Unbiased Estimator\\[1ex] for ensemble-variational data assimilation}

\ATauthor{Mayeul Destouches, Paul Mycek, Selime G\"urol}

\ATnumber{TR-PA-23-67}

\usepackage[style=authoryear-comp,
			maxcitenames=2,    
			maxbibnames=65,    
			uniquelist=false,  
			sorting=nyt,       
			sortcites=false,   
 			backend=biber,     
			natbib=true,       
			url=false,         
			urldate=long,
			uniquename=init,   
			giveninits=true,   
			isbn=false,
			backref=true, 	   
			date=year,         
			hyperref
			]{biblatex}
\inputencoding{utf8}

\makeatletter
\def\blx@maxline{77}
\makeatother

\defcitealias{schaden2020multilevel}{SU20}
\defcitealias{schaden2021asymptotic}{SU21}
\defcitealias{menetrier2015linear1}{M15a}
\defcitealias{menetrier2015linear2}{M15b}
\newcommand{\SUml}{\citetalias{schaden2020multilevel}}
\newcommand{\SUas}{\citetalias{schaden2021asymptotic}}
\newcommand{\BMa}{\citetalias{menetrier2015linear1}}
\newcommand{\BMb}{\citetalias{menetrier2015linear2}}

\newcommand{\rvline}{\hspace*{-\arraycolsep}\vline\hspace*{-\arraycolsep}}

\newcommand{\mat}[1]{\mathbf{#1}}

\newcommand{\T}{\intercal}

\newcommand{\mleft}[1]{\mathopen{}\left#1}
\newcommand{\mright}[1]{\mathclose{}\right#1}

\DeclareMathOperator{\mse}{MSE}
\DeclareMathOperator{\diag}{Diag}

\DeclareMathOperator*{\argmin}{arg\,min}

\DeclareMathOperator{\E}{\mathbb{E}}
\DeclareMathOperator{\V}{\mathbb{V}}
\DeclareMathOperator{\C}{Cov}

\newcommand{\Emc}{\ensuremath\widehat{E}}

\newcommand{\Cmc}{\ensuremath\widehat{C}}
\newcommand{\bCmc}{\ensuremath\widehat{\mathbf{C}}}

\definecolor{gplotPurple}{HTML}{9400d3}
\definecolor{gplotGreen}{HTML}{009e73}
\definecolor{gplotBlue}{HTML}{56b4e9}

\begin{document}

\begin{abstract}
Multilevel estimators aim at reducing the variance of Monte Carlo statistical estimators, by combining samples generated with simulators of different costs and accuracies.
In particular, the recent work of \citet{schaden2020multilevel} on the multilevel best linear unbiased estimator (MLBLUE) introduces a framework unifying several multilevel and multifidelity techniques.
The MLBLUE is reintroduced here using a variance minimization approach rather than the regression approach of \citeauthor{schaden2020multilevel}.
We then discuss possible extensions of the scalar MLBLUE to a multidimensional setting, i.e.\ from the expectation of \emph{scalar} random variables to the expectation of random \emph{vectors}.
Several estimators of increasing complexity are proposed: a) multilevel estimators with scalar weights, b) with element-wise weights, c) with spectral weights and d) with general matrix weights.
The computational cost of each method is discussed.
We finally extend the MLBLUE to the estimation of second-order moments in the multidimensional case, i.e.\ to the estimation of covariance matrices. 
The multilevel estimators proposed are d) a multilevel estimator with scalar weights and e) with element-wise weights.
In large-dimension applications such as data assimilation for geosciences, the latter estimator is computationnally unaffordable. 
As a remedy, we also propose f) a multilevel covariance matrix estimator with optimal multilevel localization, inspired by the optimal localization theory of \citet{menetrier2015optimized}.
Some practical details on weighted MLMC estimators of covariance matrices are given in appendix.
\end{abstract}

\tableofcontents

\section{Introduction}
Multilevel techniques aim at reducing the variance of Monte Carlo statistical estimators, typically for the estimation of the expectation of a scalar random variable.
These techniques combine in an astute way samples obtained through numerical simulators of varying accuracy and cost.
An example of popular multilevel technique is the Multilevel Monte Carlo (MLMC) method, popularized by \citet{giles2008multilevel,giles2015multilevel}.

Recently, an interesting unifying framework was proposed by \citet{schaden2020multilevel,schaden2021asymptotic}, herafter \SUml{} and \SUas.
Among others, the framework of \citeauthor{schaden2020multilevel} includes multilevel Monte Carlo techniques (MLMC, \citealp{giles2008multilevel}, \citeyear{giles2015multilevel} for a review), multifidelity techniques \citep{peherstorfer2018survey} and approximate control variates \citep{gorodetsky2020generalized}.
In this unified framework, some (new) estimators naturally appear as optimal, the so-called \emph{Multilevel Best Linear Unbiased Estimators}, MLBLUEs.
This framework has been complemented by \citet{croci2023multifidelity}, who propose an efficient algorithm to solve the \emph{model selection and sample allocation problem} (MOSAP) for the MLBLUE.

The present note proposes a new way to derive the MLBLUE, by building a weighted multilevel estimator and optimizing its weights to minimize the estimator's variance under a no-bias constraint. 
It also gives some insight on how the MLBLUE approach can be extended to the estimation of first and second-order statistical moments of random vectors, in possibly large dimensions.

This extension to second-order moments and to random vectors is motivated by possible applications in ensemble-variational data assimilation, where Monte Carlo methods are used at a key stage, to estimate the covariance matrix of forecast errors (\citealp{lorenc2003potential,buehner2005ensemble} and \citealp{bannister2017review} for a review).
As a result, the present note is not as general as the original articles by \SUml{} and \SUas, nor as mathematically grounded.
The authors are biased towards MLMC-like applications, and towards the estimation of discrete covariance operators in large dimension for geoscience applications.

The note is organized as follows.
Section \ref{sec:reminders} presents the main results of \SUml{} and \SUas{}.
Section \ref{sec:scalar-expectation} presents another way to derive these results, based on direct minimization of the variance of a weighted multilevel estimator.
The next sections proposes extensions of the MLBLUE, some of which are unpublished in the literature to the best of our knowledge.
We propose an extension to the multidimensional case (estimation of the expectation of a random vector) in section \ref{sec:expectation_random_vector}.
We propose an extension to the estimation of covariance and covariance matrices in sections \ref{sec:cov-scalar} and \ref{sec:cov-nd}, including an extension to optimal localization for multilevel covariance matrices in the line of \citet{menetrier2015linear1,menetrier2015linear2}, hereafter \BMa{} and \BMb.

\clearpage
\section{The MLBLUE: reminder and notations}
\label{sec:reminders}
Let $Z_\ell=f_\ell(X) \colon \Omega\to\mathbb{R},\, 1\leq\ell\leq L$ be a set of random variables approximating $f_L(X)$, where $f_L$ is a costly numerical simulator.
The $\ell$ indexing the $f_\ell$ models are hereafter called fidelity levels.
These fidelity levels may be associated to different spatial meshes, from the coarsest to the finest ($\ell=1$ to $\ell=L$ for an MLMC-like structure).
This is not required though, and what follows can be applied even if the fidelities come from other sources, and even if there is no clear ranking of the models according to their accuracy.

\paragraph{Example}
The hierarchy of simulators can be forecasting models $f_\ell \colon \mathbb{R}^n \to \mathbb{R}$ running on meshes with finer and finer horizontal resolutions, and predicting temperature at one given location.
$X \colon \Omega \to \mathbb{R}^n$ can be a random vector representing uncertain initial conditions of a numerical weather forecast.
We are interested in the mean temperature that is forecast by the finest model, $\E\bigl[f_L(X)\bigr]$. 

\paragraph{}
Multilevel techniques rely on coupled simulations, i.e.\ simulations that run at different levels using the same stochastic  input $X$.
The sets of coupled levels can be sorted in $K$ coupling groups $(S^{(k)})_{k=1}^K$.
More formally, let $(S^{(k)})_{k=1}^K$ be a family of subsets of $\lbrace 1,\dots,L\rbrace$.
We impose 
\begin{align}
	S^{(k)} \neq S^{(k')} \text{ for }k\neq k'\\
	\cup_{k=1}^K S^{(k)} = \{1,\dots,L\}.
\end{align}
We denote by $p^{(k)}$ the cardinality of $S^{(k)}$.

For $1\leq k\leq K$, $R^{(k)} \colon \mathbb{R}^L\to\mathbb{R}^{p^{(k)}}$ is the selection operator for group $S^{(k)}$ verifying $\forall x\in \mathbb{R}^L,\, R^{(k)}x = (x_\ell)_{\ell\in S^{(k)}}$.
The associated extension operator is 
$P^{(k)}:=\bigl(R^{(k)}\bigr)^\T \colon\mathbb{R}^{p^{(k)}}\to\mathbb{R}^L$.

\begin{figure}[ht]
	\centering
	\begin{tikzpicture}[scale=1.0, xshift=-.5]

		\coordinate (mlmcf) at (0,0);
		\draw (mlmcf) [thick] grid ++(3,-3);
		\draw (mlmcf)++(.5,-.5) node {$f_1$};
		\draw (mlmcf)++(1.5,-.5) node {$f_1$};
		\draw (mlmcf)++(1.5,-1.5) node {$f_2$};
		\draw (mlmcf)++(2.5,-1.5) node {$f_2$};
		\draw (mlmcf)++(2.5,-2.5) node {$f_3$};
		\draw (mlmcf)++(1.5,.5) node {a) Simulators $f_\ell$};
		\draw (mlmcf)++(-.75, -1.5) node [rotate=90] {fidelity level $\ell$};
		\draw (mlmcf)++(-.25, -.5) node {$1$};
		\draw (mlmcf)++(-.25, -1.5) node {$2$};
		\draw (mlmcf)++(-.25, -2.5) node {$3$};
		\draw (mlmcf)++(+.5, -3.5) node {$1$};
		\draw (mlmcf)++(+1.5, -3.5) node {$2$};
		\draw (mlmcf)++(+2.5, -3.5) node {$3$};
		\draw (mlmcf)++(1.5,-4) node {coupling group $k$};

		\coordinate (mlmc) at ($(mlmcf) + (5,0)$);
		\draw (mlmc) [thick] grid ++(3,-3);
		\draw (mlmc)++(.5,-.5) node {$m^{(1)}$};
		\draw (mlmc)++(1.5,-.5) node {$m^{(2)}$};
		\draw (mlmc)++(1.5,-1.5) node {$m^{(2)}$};
		\draw (mlmc)++(2.5,-1.5) node {$m^{(3)}$};
		\draw (mlmc)++(2.5,-2.5) node {$m^{(3)}$};
		\draw (mlmc)++(1.5,.5) node {b) Ensemble sizes};
		\draw (mlmc)++(-.75, -1.5) node [rotate=90] {fidelity level $\ell$};
		\draw (mlmc)++(-.25, -.5) node {$1$};
		\draw (mlmc)++(-.25, -1.5) node {$2$};
		\draw (mlmc)++(-.25, -2.5) node {$3$};
		\draw (mlmc)++(+.5, -3.5) node {$1$};
		\draw (mlmc)++(+1.5, -3.5) node {$2$};
		\draw (mlmc)++(+2.5, -3.5) node {$3$};
		\draw (mlmc)++(1.5,-4) node {coupling group $k$};

		\coordinate (mlmcbeta) at ($(mlmc) + (5,0)$);
		\draw (mlmcbeta) [thick] grid ++(3,-3);
		\draw (mlmcbeta)++(.5,-.5) node {$1$};
		\draw (mlmcbeta)++(1.5,-.5) node {$-1$};
		\draw (mlmcbeta)++(1.5,-1.5) node {$1$};
		\draw (mlmcbeta)++(2.5,-1.5) node {$-1$};
		\draw (mlmcbeta)++(2.5,-2.5) node {$1$};
		\draw (mlmcbeta)++(1.5,.5) node {c) Weights $\beta_\ell^{(k)}$};
		\draw (mlmcbeta)++(-.75, -1.5) node [rotate=90] {fidelity level $\ell$};
		\draw (mlmcbeta)++(-.25, -.5) node {$1$};
		\draw (mlmcbeta)++(-.25, -1.5) node {$2$};
		\draw (mlmcbeta)++(-.25, -2.5) node {$3$};
		\draw (mlmcbeta)++(+.5, -3.5) node {$1$};
		\draw (mlmcbeta)++(+1.5, -3.5) node {$2$};
		\draw (mlmcbeta)++(+2.5, -3.5) node {$3$};
		\draw (mlmcbeta)++(1.5,-4) node {coupling group $k$};

	\end{tikzpicture}
	\caption{MLMC coupling structure. Tableaux inspired by \SUml{}.}
	\label{fig:mlmc-ex}
\end{figure}
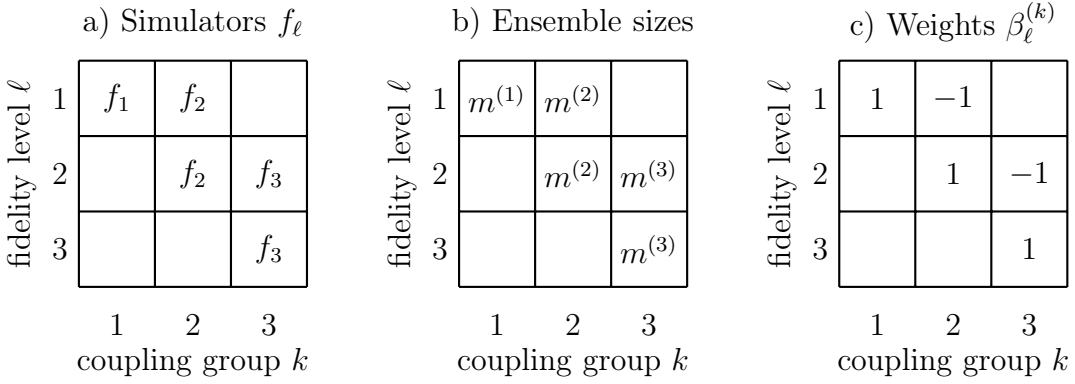

\paragraph{MLMC-like example}
We can use this formalism to describe the coupling structure of an MLMC estimator with three levels, as is done in example~2.1 of \citetalias{schaden2020multilevel}. 
The coupling groups in this case are $S^{(1)} = \{1\}$, $S^{(2)}=\{1,2\}$ and $S^{(3)}=\{2,3\}$.
The associated extension operators are
\begin{align}
	P^{(1)} &= \begin{pmatrix}1&0&0\end{pmatrix}\\
	P^{(2)} &= \begin{pmatrix}
		1&0&0\\
		0&1&0
		\end{pmatrix}\\
	P^{(3)} &= \begin{pmatrix}
		0&1&0\\
		0&0&1
	\end{pmatrix}
\end{align}
The repartition of simulators among coupling groups $S^{(k)}$ can be visualized using the tableaux used by \citeauthor{schaden2020multilevel}, and reproduced for this example in figure~\ref{fig:mlmc-ex}c.

We denote by $\mu := \left(\E\left[Z_\ell\right]\right)_{\ell=1}^L$ the vector of expectations at each fidelity level.
We are interested in estimating $\alpha^\T\mu$  for a given vector $\alpha\in\mathbb{R}^L\setminus\lbrace 0\rbrace$.
In practice, $\alpha=e_L:=(0,\cdots,0, 1)^\T$, but this is not mandatory.

Let $m^{(1)},\dots,m^{(K)}$ be the number of available simulations for each group $k$ (see figure \ref{fig:mlmc-ex}b).
\SUml{} provide the best estimator for $\alpha^\T\mu$ among the unbiased estimators that linearly combine the simulations $f_\ell(X^{(k,i)})$ for $1\leq k\leq K$, $\ell\in S^{(k)}$ and $1\leq i \leq m^{(k)}$, where the $X^{(k,i)}$ are i.i.d.\ random variables following the same law as $X$.
These linear estimators are of the form

\begin{align}
\widehat{\mu}^\text{ML} := \sum_{k=1}^K\sum_{\ell\in S^{(k)}} \beta_{\ell}^{(k)} \Emc^{(k)}[Z_\ell]\label{eq:coeff_ml}
\end{align}
where $\Emc^{(k)}[Z_\ell]$ is the standard Monte Carlo estimator for $\E[Z_\ell]$, using $m^{(k)}$ random inputs associated to group $k$,
\begin{align}
\Emc^{(k)}[Z_\ell] := \frac{1}{m^{(k)}} \sum_{i=1}^{m^{(k)}}f_\ell(X^{(k,i)}).
\end{align}

The inner sum in \eqref{eq:coeff_ml} can be written as a scalar product by introducing two more notations.
Firstly, we denote by $\beta^{(k)}:=\bigl(\beta_{\ell}^{(k)}\bigr)_{\ell\in S^{(k)}}\in\mathbb{R}^{p^{(k)}}$ the vector of weights associated to group $k$.
In the case of a three-level MLMC, we would have the (sub-optimal) weights $\beta^{(1)}=(1)$ and $\beta^{(2)} = \beta^{(3)} = (-1, 1)^\T$ (figure \ref{fig:mlmc-ex}c).
Secondly, we denote by $Z^{(k)} := (Z_\ell)_{\ell\in S^{(k)}}$ the random vector gathering all random variables in group $k$.
Then equation \eqref{eq:coeff_ml} becomes
\begin{align}
\widehat{\mu}^\text{ML}&= \sum_{k=1}^K \left(\beta^{(k)}\right)^\T \Emc^{(k)}\bigl[Z^{(k)}\bigr]\label{eq:coeff_ml_betak}
\end{align}

The (optimal) vector $\beta^{(k)}$ can be expressed as (from equation 2.7 in \SUas)
\begin{align}
\beta^{(k)} = m^{(k)}\left(C^{(k)}\right)^{-1}R^{(k)}\left(\sum_{k'=1}^K m^{(k')} P^{(k')}\left(C^{(k')}\right)^{-1}R^{(k')}\right)^{-1}\alpha,
\end{align}
where $C^{(k)}:=\C(Z^{(k)}, Z^{(k)})$ and $\C(A, B) := \E[\left(A-\E[A]\right)\left(B-\E[B]\right)^\T]$ denotes the covariance matrix of random vectors $A$ and $B$.
These $C^{(k)}$ matrices are unknown in practice and must be estimated, which results in sub-optimal weights.
Note that the estimation of the $C^{(k)}$ should be done independently of the estimation of $\alpha^\T \mu$, otherwise a bias is introduced.
The importance of this bias is likely to depend on the particular application and setting considered.

\paragraph{Model selection and sample allocation problem}
This approach provides the MLBLUE for a given coupling structure and a given number of samples on each coupling group. 
\citetalias{schaden2020multilevel} propose some ways to optimize the model selection and sample allocation by minimizing the variance of the associated MLBLUE. 
Their approach has been later extended and made more robust by \citet{croci2023multifidelity}, who transform it into a semidefinite programming problem.

\clearpage
\section{Retrieving the MLBLUE via variance minimization}
\label{sec:scalar-expectation}
We believe the derivation based on constrained minimization of the variance to be more direct, and perhaps more intuitive than the regression approach proposed by \SUml{}.
Though both derivations are closely related, we believe the variance minimization approach may appear as more natural to some readers, especially from the community of multifidelity estimation methods based on control variates \citep[see for instance][]{gorodetsky2020generalized}.
We derive this approach here.

Given the fidelity levels $1,\ldots, L$ and the coupling structure $(S^{(k)}, m^{(k)})_{k=1}^K$, we look for an unbiased estimator of $\alpha^\T \mu$ that linearly combines the samples and that has the lowest possible variance (the BLUE).
We assume that the multilevel estimator is of the form of equations \eqref{eq:coeff_ml} and \eqref{eq:coeff_ml_betak}, and we look for the $\beta_\ell^{(k)}$ coefficients that minimize the variance under a no-bias constraint. 

\paragraph{Unbiasedness constraint}
The optimal $\beta$ weights are subject to the unbiasedness constraint
\begin{align}
\E\mathopen{}\left[\widehat{\mu}(\beta)\right]\mathclose{} = \alpha^\T \mu \quad 
&\Longleftrightarrow\quad\E\mathopen{}\Biggl[\sum_{k=1}^K \left(\beta^{(k)}\right)^\T \Emc^{(k)}\bigl[Z^{(k)}\bigr]\Biggr] = \alpha^\T \mu\\
&\Longleftrightarrow\quad\sum_{k=1}^K \left(\beta^{(k)}\right)^\T R^{(k)}\mu = \alpha^\T \mu\\
&\Longleftrightarrow\quad\sum_{k=1}^K P^{(k)} \beta^{(k)} = \alpha, \label{eq:no-bias-tricky-part}
\end{align}
assuming that we have no prior information on $\mu$, so that the unbiasedness should be met for all values of $\mu\in\mathbb{R}^L$.
\begin{align}
	\E\mathopen{}\left[\widehat{\mu}(\beta)\right]\mathclose{} = \alpha^\T \mu \quad 
	&\Longleftrightarrow\quad \begin{pmatrix}P^{(1)} &\cdots & P^{(K)}\end{pmatrix} \beta = \alpha\label{eq:no-bias}\\
	&\Longleftrightarrow\quad g(\beta) = 0\label{eq:no-bias-g}\\
	&\text{with } g(\beta) := \begin{pmatrix}P^{(1)} &\cdots & P^{(K)}\end{pmatrix} \beta - \alpha\label{eq:def_of_g}
\end{align}
and where we denote by $\beta := \left(\beta^{(k)}\right)_{k=1}^K\in\mathbb{R}^p$ the vector made of all the $\beta^{(k)}$, with $p:=\sum_{k=1}^K p^{(k)}$\label{def:p}.

\paragraph{Expression of the variance}

The variance of the linear estimator $\widehat{\mu}$ is the sum of the variances for each coupling group, since simulations are independent from one coupling group to another.
We denote by $\V(X)$ the variance of any square-integrable random variable $X$.
\begin{align}
\V(\widehat{\mu}(\beta)) &= \sum_{k=1}^K  \V\mleft(\left(\beta^{(k)}\right)^\T \Emc^{(k)}\bigl[Z^{(k)}\bigr]\mright)\\
&=\sum_{k=1}^K  \left(\beta^{(k)}\right)^\T\C\mleft(\Emc^{(k)}\bigl[ Z^{(k)}\bigr], \Emc^{(k)}\bigl[ Z^{(k)}\bigr]\mright)\beta^{(k)}\label{eq:before_indep_i}\\
&=\sum_{k=1}^K  \frac{1}{m^{(k)}}\left(\beta^{(k)}\right)^\T\C\mleft( Z^{(k)}, Z^{(k)}\mright)\beta^{(k)}\label{eq:after_indep_i}\\
&=\sum_{k=1}^K  \frac{1}{m^{(k)}} \left(\beta^{(k)}\right)^\T C^{(k)}\beta^{(k)}\label{eq:2.8}\\
&= \beta^\T \mathit{\Sigma} \beta\\
\text{with }  \mathit{\Sigma} &:= \diag_{k=1}^K\mleft(\frac{1}{m^{(k)}}C^{(k)}\mright).
\end{align}
We used the independence of the $f_\ell\mleft(X^{(k, i)}\mright)$ for different $i$ to go from \eqref{eq:before_indep_i} to \eqref{eq:after_indep_i}, and used the $\diag$ operator to denote a block-diagonal matrix.
Equation \eqref{eq:2.8} is equivalent to equation (2.8) in \SUas.

\paragraph{Convexity of the variance}
$\mathit{\Sigma}$ is a block-diagonal matrix with covariance matrices on the diagonal.
As a result, it is positive semi-definite and $\V(\widehat{\mu}(\beta))$ is a convex (quadratic) function of $\beta$.
Note that as will be discussed hereafter, the covariance matrices $C^{(k)}$ are actually positive definite, and the variance is a strictly convex function of $\beta$. 

\paragraph{Constrained minimization problem}
The best unbiased estimator is then given by the minimizer of the variance under the unbiasedness constraint.

\begin{equation}
  \beta^\star
  =
  \argmin_{\beta\text{ s.t.\ } g(\beta) = 0}
  \frac{1}{2}\V\bigl(\widehat{\mu}(\beta)\bigr).\label{eq:constrained_problem}
\end{equation}

\paragraph{Unconstrained minimization problem}
The assumptions on the coupling groups $S^{(k)}$ ensure that the $L$ constraints are linearly independent.
A vector $\lambda\in\mathbb{R}^L$ of Lagrange multipliers can be used to solve the minimization problem.
From the convexity of the quadratic problem, the solutions of \eqref{eq:constrained_problem} are the solutions of
\begin{align}
\beta^\star, \lambda^\star &= \argmin_{\beta, \lambda} \mathcal{L}(\beta, \lambda),\\
\text{with } \mathcal{L}(\beta, \lambda) &=\frac{1}{2}\V(\widehat{\mu}(\beta)) - \lambda^\T g(\beta).
\end{align}
In particular, the gradient of the Lagrangian,
\begin{align}
\nabla_\beta \mathcal{L} &= \beta^\T \operatorname{Diag}_{k=1}^K\mleft(\frac{1}{m^{(k)}}C^{(k)}\mright) - \lambda^\T \begin{pmatrix}P^{(1)} &\cdots & P^{(K)}\end{pmatrix},\\
\nabla_\lambda \mathcal{L} &= \beta^\T\begin{pmatrix}P^{(1)} &\cdots & P^{(K)}\end{pmatrix}^\T - \alpha^\T,
\end{align}
should vanish. The associated linear system is
\begin{align}
\begin{pmatrix}
\frac{1}{m^{(1)}}C^{(1)}& & &\rvline & -R^{(1)}\\
&\ddots & &\rvline& \vdots\\
& & \frac{1}{m^{(k)}}C^{(K)} &\rvline& -R^{(K)}\\[1ex]
\hline\rule{0em}{2.5ex}
P^{(1)}&\cdots & P^{(K)} &\rvline & 0_L
\end{pmatrix}
\begin{pmatrix}
\\[1ex]
\beta^\star\\
\\[1ex]
\hline\rule{0em}{2.5ex}
\lambda^\star
\end{pmatrix}
=
\begin{pmatrix}
0\\
\vdots\\
0\\[1ex]
\hline\rule{0em}{2.5ex}
\alpha
\end{pmatrix}
\end{align}
To simplify the notations, we drop the stars and write the system as
\begin{align}
\mathit{\Sigma}\beta -P^\T\lambda &= 0,\label{eq:sys1}\\
P\beta &= \alpha. \label{eq:sys2}
\end{align}
This system can be solved by substitution:
\begin{align}
\text{From  \eqref{eq:sys1}:}& &\beta &= \mathit{\Sigma}^{-1}P^\T\lambda\label{eq:sys3}\\
\text{Inserting \eqref{eq:sys3} in \eqref{eq:sys2}:}& & P\mathit{\Sigma}^{-1}P^\T\lambda &= \alpha\label{eq:sys4}\\
& &\iff \lambda &= \left(P\mathit{\Sigma}^{-1}P^\T\right)^{-1}\alpha\label{eq:sys5}\\
\text{Inserting \eqref{eq:sys5} in \eqref{eq:sys3}:}& &\beta &= \mathit{\Sigma}^{-1}P^\T\left(P\mathit{\Sigma}^{-1}P^\T\right)^{-1}\alpha. \label{eq:sys-last}
\end{align}

\paragraph{Invertibility of the matrices}
We assumed the invertibility of $\mathit{\Sigma}$ and $P\mathit{\Sigma}^{-1}P^\T$.
The invertibility of $\mathit{\Sigma}$ follows from the invertibility of each covariance matrix $C^{(k)}$.
Suppose $\mathit{\Sigma}$ is singular.
Then, there exists a $k$ such that $C^{(k)}$ is singular.
Then there exists a vector $\gamma\in\mathbb{R}^{p^{(k)}}\setminus\{0\}$ such that $\gamma^\T C^{(k)} \gamma = \V(\gamma^\T Z^{(k)}) = 0$, \emph{i.e.} $\sum_{\ell\in S^{(k)}}\gamma_\ell Z_\ell$ is actually deterministic.
One random variable $Z_\ell$ can thus be expressed as an affine function of the others.
It brings no new information to the problem, and can be removed from the estimator.
The invertibility of $P\mathit{\Sigma}^{-1}P^\T$ follows from the invertibility of $\mathit{\Sigma}^{-1}$ and from $P$ being a full-rank linear map from $\mathbb{R}^p$ to the lower-dimensional space $\mathbb{R}^L$.

\paragraph{MLBLUE weights}
The optimal choice of $\beta$ is thus given by equation \eqref{eq:sys-last}:
\begin{align}
\beta &=
\begin{pmatrix}
\beta^{(1)}\\ \vdots \\ \beta^{(K)}	
\end{pmatrix} = 
\begin{pmatrix}
m_1\left(C^{(1)}\right)^{-1}& &\\
&\ddots & \\
& & m^{(k)}\left(C^{(K)}\right)^{-1}\\
\end{pmatrix}
\begin{pmatrix} R^{(1)}\\ \vdots \\ R^K \end{pmatrix}
\phi^{-1}\alpha,
\end{align}
where
\begin{align}
\phi := P\Sigma^{-1}P^\T = \sum_{k=1}^K m^{(k)} P^{(k)}\left(C^{(k)}\right)^{-1}R^{(k)}.
\end{align}
For a given group $k$, we retrieve equation (2.7) of \SUas, namely
\begin{align}
\beta^{(k)} = m^{(k)} \left(C^{(k)}\right)^{-1}R^{(k)} \left(\sum_{k'=1}^K m^{(k')} P^{(k')}\left(C^{(k')}\right)^{-1}R^{(k')}\right)^{-1}\alpha \label{eq:2.7}.
\end{align}

\paragraph{Is it the MLBLUE?} The estimators of the form \eqref{eq:coeff_ml_betak} only describe a specific subset of all possible linear estimators.
The more general class of linear estimators would be
\begin{align}
\mu^\text{ML} = \sum_{k=1}^K\sum_{\ell\in S^{(k)}}\sum_{i=1}^{m^{(k)}}\beta_\ell^{(k,i)}f_\ell\mleft(X^{(k,i)}\mright).\label{eq:blue_without_interchangeability}
\end{align}
Estimators of the form \eqref{eq:blue_without_interchangeability} can be related to Eq.~\eqref{eq:coeff_ml_betak} by replacing $\beta_\ell^{(k,i)}$ with $\beta_\ell^{(k)}/m^{(k)}$.
In other words, Eq.~\eqref{eq:coeff_ml_betak} assumes that at the optimum, the $\beta_\ell^{(k,i)}$ weights should not depend on the sample index $i$.
This independence is a very intuitive result, that can be derived from the interchangeability of the samples $i$ and the strict convexity of the variance of \eqref{eq:blue_without_interchangeability} as a function of the weights.

\paragraph{Sample allocation}
Inserting \eqref{eq:2.7} into \eqref{eq:2.8} and simplifying gives the expression of the minimum variance reachable with a given sample allocation $m$ (equivalent to equation 2.12 in \SUml):
\begin{align}
	\V(\mu^\text{ML}(m)) = \alpha^\T\phi(m)^{-1}\alpha\label{eq:optvar_scalar}.
\end{align}
The optimal choice for the sample allocation $m=(m^{(1)}, \dots, m^{(K)})$ is given by minimizing this variance under a computational cost constraint, which can be done numerically.

\paragraph{Model selection and sample allocation problem}
Alternatively, the model selection and sample allocation problem (MOSAP) can be solved through a semidefinite programming problem, as shown by \citet{croci2023multifidelity}, in the typical case where $\alpha = e_L$:

\begin{equation}
  \min\limits_{m\geq 0, \,t} t \quad \text{s.t.} \quad 
  \begin{cases}
    \begin{pmatrix}
      \phi(m) & e_L\\
      e_L^\T & t
    \end{pmatrix} \text{ is positive semi-definite},\\
    m^\T c \leq b, \\
    m^\T h \geq 1.
  \end{cases}\label{eq:mosap}
\end{equation}
The second constraint imposes a computational budget $b$, where $c=(c^{(1)},\dots,c^{(K)})^\T$ describes the computational cost of generating a coupled sample in each coupling group. 
The third constraint, where $h$ denotes the vector of $\{0;1\}^K$ such that $h^{(k)} =1$ if and only if $L\in S^{(k)}$, enforces that the high-fidelity model be sampled at least once.

Note that this extends easily to the case of any $\alpha$ by replacing $e_L$ with $\alpha$ and by adding inequality constraints to ensure that all model with non-zero coefficients in $\alpha$ are sampled at least once. 
Also note that this approach to the MOSAP can handle sample sizes of zero, which means the problem can be directly optimized on the set of all possible coupling groups. 
This would not be directly possible with the sample allocation strategy proposed previously.

A similar version for a target accuracy with no constraint on the computational budget is also proposed in \citet{croci2023multifidelity}.

\clearpage
\section{Estimation of the expectation of a random vector}
\label{sec:expectation_random_vector}

This section extends the MLBLUE methodology to the estimation of the expectation of a random vector.
Section (\ref{sec:nd-notations}) introduces new notations to deal with vector quantities.
We then propose various multilevel estimators of increasing complexity (sections \ref{sec:nd-scalar} -- \ref{sec:nd-Wfield}), before introducing the general multidimensional MLBLUE in section \ref{sec:nd-fullmatrixweights}.

\subsection{Notations for the multidimensional case}
\label{sec:nd-notations}
We consider the case where $\mathbf{Z}_\ell:\,\Omega\to\mathbb{R}^n$ are random vectors.
All vectors or matrices related to multidimensional quantities are written in bold.

We are interested in the expectation $\bm{\mu} := \E\bigl[\mathbf{Z}\bigr]$, where $\mathbf{Z} := (\mathbf{Z}_1 \dots \mathbf{Z}_L)^\T$ is the random matrix with values in $\mathbb{R}^{L\times n}$, obtained by stacking the random vectors from all fidelity levels so that the first dimension of $\mathbf{Z}$ indexes the fidelity levels.

We want to estimate a linear combination of the expectations on different levels, $\bm{\mu}_\alpha := \sum_{\ell=1}^L \alpha_\ell \E\bigl[\mathbf{Z}_\ell\bigr] = \bm{\mu}^\T\alpha$, where $\alpha$ is a non-zero vector of $\mathbb{R}^L$.
In practice, we are often interested in the estimation for one given fidelity level, typically the highest, in which case $\alpha=e_L$.

The selection and extension operators from the scalar case naturally extend to the vector case:
\begin{align}
\mathbf{Z}^{(k)} := R^{(k)}\mathbf{Z}\in\mathbb{R}^{p^{(k)}\times n}, \quad\forall\, 1\leq k\leq K.
\end{align}

\paragraph{Vector equivalent of the variance}
Under a no-bias constraint, the variance of a random variable is its (scalar) mean squared error (MSE). 
In the multidimensional case, minimizing the MSE of an estimator $\widehat{\bm{\mu}}$ using the 2-norm is equivalent to minimizing the sum of scalar MSEs for all vector elements.
\begin{align}
\E\bigl[\|\widehat{\bm{\mu}} -  \bm{\mu} \|_2^2\bigr] &= \sum_{i=1}^n \E\bigl[\left(\widehat{\mu}_i -  \mu_{i}\right)^2\bigr]\\
&= \sum_{i=1}^n \V(\widehat{\mu}_i)\label{eq:var_mul_nd}\\
&= \operatorname{Tr}\C\left(\widehat{\bm{\mu}},\widehat{\bm{\mu}}\right)
\end{align}
where $\operatorname{Tr}$ is the trace operator, and where we used the unbiasedness of the estimator $\widehat{\bm{\mu}}$. 
It can be seen from here that the natural generalization of the variance for a random vector is the trace of the covariance matrix, \emph{i.e.} the sum of the variances of each vector element. 

Each of the following sections introduces a class of estimators $\widehat{\bm{\mu}}(\bm{\beta})$  where $\bm{\beta}$ is a set of weights. 
Similarly to the scalar case, the optimal value $\bm{\beta}^\star$ of these weights is found by minimization of the trace of the covariance matrix of the estimator, under a no-bias constraint:
\begin{equation}
	\bm{\beta}^\star
	=
	\argmin_{\bm{\beta}\text{ s.t.\ } \mathbb{E}[\widehat{\bm{\mu}}(\bm{\beta})] = 0}
	\operatorname{Tr}\C\bigl(\widehat{\bm{\mu}}(\bm{\beta}), \widehat{\bm{\mu}}(\bm{\beta})\bigr).
\end{equation}
Hereafter, the term \emph{variance} is sometimes used to refer to the trace of the covariance matrix.

\subsection{Scalar weights}
\label{sec:nd-scalar}
The simplest possibility is to use scalar weights $\beta_\ell^{(k)}$, common to all random vector elements.
In this case, $\bm{\mu}_\alpha$ is estimated through the linear combination of Monte Carlo estimators
\begin{align}
\widehat{\bm{\mu}}_\alpha^{\text{sw}} &= \sum_{k=1}^K\sum_{\ell\in S^{(k)}} \beta_\ell^{(k)} \Emc^{(k)}\bigl[\mathbf{Z}_\ell\bigr]\\
&= \sum_{k=1}^K \Emc^{(k)}\bigl[\mathbf{Z}^{(k)}\bigr]^\T\beta^{(k)}.
\end{align}
The BLUE has no reason to lie in this class of estimators, which is just a subset of the linear estimators we will introduce in sections \ref{sec:nd-field} to \ref{sec:nd-fullmatrixweights}.
Finding the optimal scalar weights $\beta_\ell^{(k)}$ is still of interest though, as these scalar-weigthed estimators are both simple and numerically affordable. 

\paragraph{Unbiasedness constraint}
\begin{align}
	&&\E\bigl[\widehat{\bm{\mu}}_\alpha^{\text{sw}}\bigr] &= \bm{\mu}_\alpha\rule{9em}{0ex}\\
	&\Longleftrightarrow& \E\Biggl[\sum_{k=1}^K \Emc^{(k)}\bigl[\mathbf{Z}^{(k)}\bigr]^\T\beta^{(k)}\Biggr] &= \E[\mathbf{Z}^\T]\alpha\\
	&\Longleftrightarrow& \sum_{k=1}^K \E\bigl[\mathbf{Z}^{(k)}\bigr]^\T\beta^{(k)} &= \E[\mathbf{Z}^\T]\alpha\\
	&\Longleftrightarrow& \sum_{k=1}^K \E\bigl[R^{(k)}\mathbf{Z}\bigr]^\T\beta^{(k)} &= \E[\mathbf{Z}]^\T\alpha\\
	&\Longleftrightarrow& \E\bigl[\mathbf{Z}\bigr]^\T\sum_{k=1}^K P^{(k)}\beta^{(k)} &= \E[\mathbf{Z}]^\T\alpha\\
	&\Longleftrightarrow& \sum_{k=1}^K P^{(k)} \beta^{(k)} &= \alpha\\
	&\Longleftrightarrow& g(\beta) &= 0\label{eq:nobias_sc}
\end{align}
where $g$ was defined in Eq. \eqref{eq:def_of_g}. 
This is exactly the unbiasedness constraint \eqref{eq:no-bias-g} from the scalar expectation case. 

\paragraph{Variance of the estimator}
Herafter, the covariance operator is occasionally extended to random matrices using the definition
\begin{align}
	\C\mathopen{}\bigl(\mathbf{A}, \mathbf{B}\bigr) := \E\mathopen{}\bigl[\bigl(\mathbf{A} - \E\mathopen{}[\mathbf{A}]\bigr)\bigl(\mathbf{B} - \E\mathopen{}[\mathbf{B}]\bigr)^\T\bigr]
\end{align}
where $\mathbf{A}$ and $\mathbf{B}$ are any random matrices with same number of columns. 
This definition coincides with the usual one in the case of column vectors. 

\begin{align}
	\operatorname{Tr}\C\left(\widehat{\bm{\mu}}_\alpha^{\text{sw}}, \widehat{\bm{\mu}}_\alpha^{\text{sw}}\right)	&= \sum_{k=1}^K \frac{1}{m^{(k)}} \operatorname*{Tr}\C \bigl( \mathbf{Z}^{(k)\T}\beta^{(k)} ,\, \mathbf{Z}^{(k)\T}\beta^{(k)}\bigr)\\
	&=\sum_{k=1}^K \frac{1}{m^{(k)}} \operatorname*{Tr}\C \bigl( \beta^{(k)\T} \mathbf{Z}^{(k)} ,\, \beta^{(k)\T} \mathbf{Z}^{(k)}\bigr)\\
	&=\sum_{k=1}^K \frac{1}{m^{(k)}} \operatorname*{Tr}\bigl\{\beta^{(k)\T} \C \bigl(\mathbf{Z}^{(k)} , \mathbf{Z}^{(k)}\bigr)\beta^{(k)}\bigr\}\\
	&=\sum_{k=1}^K \frac{1}{m^{(k)}} \beta^{(k)\T} \C \bigl(\mathbf{Z}^{(k)} , \mathbf{Z}^{(k)}\bigr)\beta^{(k)}\\
	&=\sum_{k=1}^K \frac{1}{m^{(k)}} \beta^{(k)\T} \overline{C^{(k)}}\beta^{(k)}\label{eq:var_sc}
\end{align}
with $\overline{C^{(k)}} := \C \bigl(\mathbf{Z}^{(k)} , \mathbf{Z}^{(k)}\bigr)$. 
Note that the variance of the estimator has the same expression as in the scalar expectation case (cf. Eq. \ref{eq:2.8}), just replacing $C^{(k)}$ with $\overline{C^{(k)}}$. 

\paragraph{Interpretation as averaged covariance matrices}
The relation $C^{(k)} = R^{(k)} C P^{(k)}$ that defines the $C^{(k)}$ as submatrices of the inter-level covariance matrix is still valid:
\begin{align}
	\overline{C^{(k)}} = R^{(k)}\overline{C}P^{(k)}\\
	\text{with  } \overline{C} := \C \bigl(\mathbf{Z} , \mathbf{Z}\bigr)
\end{align}
The matrix $\overline{C}$ is just the average of the inter-level covariance matrices that would be estimated for each element of the random vectors.
\begin{align}
	\overline{C} &= \sum_{i=1}^n C^{(k, i)}\\
	\text{where }\;C^{(k,i)}&:= \C((\mathbf{Z}_{:, i}), (\mathbf{Z}_{:, i}))
\end{align}
and where $\mathbf{Z}_{:, i}$ is the $i$-th column of $\mathbf{Z}$.

\paragraph{Optimal weights and MOSAP}
From previous paragraphs, it is clear that all results of the scalar expectation case now apply, just replacing the inter-level covariances with averaged inter-level covariances.

For instance, Eq.\eqref{eq:2.7} from the scalar case becomes
\begin{align}
\beta^{(k)} &= m^{(k)}\left(\overline{C^{(k)}}\right)^{-1}R^{(k)}\phi^{-1}\alpha,\label{eq:beta_nd}\\
\text{with } \phi &= \sum_{k=1}^K m^{(k)} P^{(k)}\left(\overline{C^{(k)}}\right)^{-1}R^{(k)}
\end{align}
and the MOSAP can be solved by updating $\phi(m)$ in Eq. \eqref{eq:mosap}.

\paragraph{Application in large dimensions}
This approach is tractable for large dimension systems.
The most expensive steps have a computational cost that is linear in $n$:
\begin{itemize}
	\item For each of the $n$ grid points (or vector elements more generally), estimate an $L\times L$ covariance matrix. 
	Then, take the average of these $n$ matrices.
	This averaging step should help reducing the sampling noise in the estimation of the covariance matrices. 
	\item As in the scalar expectation case, inverting a few matrices of size at most $L\times L$. 
\end{itemize}

\subsection{Field weights}
\label{sec:nd-field}
A more refined approximation consists in allowing for different $\beta$ weights depending on the element consider.
This has been introduced by \citet{croci2023multifidelity} as a ``multi-output'' MLBLUE. 
We don't propose anything new in this section compared to their work.

When the random vector to be estimated can be considered as a discretized random field, the weights $\beta$ of the multi-output MLBLUE are varying in space.
Here, we call this estimator the $\beta$-field multilevel estimator:
\begin{align}
\widehat{\bm{\mu}}^{\text{fw}}_\alpha = \sum_{k=1}^K\sum_{\ell\in S^{(k)}} \operatorname{Diag}\Bigl(\bm{\beta}_\ell^{(k)}\Bigr) \Emc^{(k)}\bigl[\mathbf{Z}_{\ell}\bigr]
\end{align}
where the $\bm{\beta}_{\ell}^{(k)}$ are now vectors of $\mathbb{R}^n$ and $\operatorname{Diag}\Bigl(\bm{\beta}_\ell^{(k)}\Bigr)$ is the diagonal matrix with diagonal $\bm{\beta}_{\ell}^{(k)}$.

This class of multilevel estimators includes the previous one (scalar weights).
This implies that the optimal $\beta$-field estimator is at least as good as the estimator with scalar weights.
Nonetheless, it is still a specific class of estimators, that has no reason to include the BLUE.

The minimization problem in this case consists in minimizing a sum of independent problems similar to \eqref{eq:constrained_problem}.
As a consequence, for a given $i\in\lbrace 1,\dots, n\rbrace$, the optimal weights $\beta_{\ell,i}^{(k)}$ are the MLBLUE weights for the scalar random variables $(Z_{\ell, i})_{\ell=1}^L$.

\paragraph{Mean value of space-dependent weights}
Note that the mean values of the $\beta_{\ell}^{(k)}$ vectors differ from the optimal scalar weights \eqref{eq:beta_nd}, due to the non-linearity introduced by the inverse in \eqref{eq:beta_nd}.

\paragraph{Variance of the estimator} 
The variance of the $\beta$-field ML estimator is given by

\begin{align}
\V\left(\widehat{\bm{\mu}}^{\text{fw}}_\alpha(m)\right) &= \sum_{i=1}^n \alpha^\T \phi_i(m)^{-1}\alpha\\
\text{where }\phi_i(m) &:= \sum_{k=1}^K m^{(k)}P^{(k)}\left(C^{(k,i)}\right)^{-1}R^{(k)}.
\end{align}

\paragraph{MOSAP}
\citet{croci2023multifidelity} proposes a solution that minimizes the maximum variance (their equation 21). 
They also propose variants to minimize, for instance, the total variance:
\begin{align}
	\min\limits_{m\geq 0, \,\mathbf{t}\in\mathbb{R}^n} \|\mathbf{t}\|_1\quad \text{s.t. } 
	\left\lbrace
	\begin{matrix}
		&\begin{pmatrix}
			\phi_i(m) & \alpha\\
			\alpha^\T & t_i
		\end{pmatrix} \text{ is positive semi-definite },  \forall i=1,\dots,n\\
		&m^\T c \leq b \\
		&m^\T h \geq 1
	\end{matrix}
	\right .
\end{align}

\paragraph{Application in large dimensions}
Finding the optimal field weights does not require much more computations than the scalar weight approach. 
Both require the estimation of the $C^{(k,i)}$ matrices.
The $\beta$-field approach requires to store $L(L+1)/2$ vectors of size $n$ to store the local covariance matrices.
More importantly, it also requires $pn$ inversions of $K$ matrices of size less than $\max_k p^{(k)}$ and of one matrix of size $L$.

\subsection{Field weights with change of basis}
\label{sec:nd-Wfield}
\paragraph{Rationale}
Another still larger (but still suboptimal) class of ML estimator can be defined by applying the estimator in a possibly different space.
This approach is motivated by the intuition that a better variance reduction could be obtained with the field-weight estimator if the data could be linearly transformed into a space where the elements are distributed according to the strength of their interlevel coupling. 

For instance, if the low-fidelity models originate from coarse grid simulations, a scale decomposition could provide such a transform, with loose coupling on fine scales and strong coupling on large scales.
In this case, there would be a set of $\beta_{\ell}^{(k)}$ weights for each wave number instead of each vector element.

This class of estimator is especially interesting for two reasons:
\begin{enumerate}
	\item It is the largest class of estimators that are still computationally tractable in high dimension, i.e. with a computational cost in $\mathcal{O}(n\log(n))$ with respect to the vector size $n$.
	\item It can be interpreted as an optimal post-smoothing, similar to what is done in multigrid methods.
\end{enumerate}
The derivations here may be a bit cumbersome though, so the impatient reader is invited to go directly to section \ref{sec:nd-fullmatrixweights} on the general multivariate MLBLUE.

\paragraph{Definition}
Let $\mathbf{W}\in\mathbb{R}^{n\times n}$ be an orthonormal matrix, so that $\mathbf{W}^\T\mathbf{W} = \mathbf{W}\mathbf{W}^\T = \mathbf{I}_n$.
We introduce the class of $W$-field multilevel estimators:
\begin{align}
\widehat{\bm{\mu}}^{\text{W}}_\alpha &:= \sum_{k=1}^K\sum_{\ell\in S^{(k)}} \mathbf{W}\operatorname{Diag}\Bigl(\bm{\beta}_\ell^{(k)}\Bigr)\mathbf{W}^\T \Emc^{(k)}\bigl[\mathbf{Z}_{\ell}\bigr]\\
 &= \mathbf{W}\sum_{k=1}^K\sum_{\ell\in S^{(k)}} \operatorname{Diag}\bigl(\bm{\beta}_\ell^{(k)}\bigr)\Emc^{(k)}\left[\mathbf{W}^\T \mathbf{Z}_{\ell}\right].
\label{eq:def_Wfield}
\end{align}

\paragraph{Relation to the $\beta$-field estimators}

\emph{
Intuitively, the optimal $W$-field estimator should be obtained by applying the optimal $\beta$-field estimator to the transformed samples $\mathbf{W}^\T\mathbf{Z}_\ell$, and transforming the result back to the physical space.
This intuitive result is now properly derived.}\\

To simplify the notations, let us denote by $\widehat{\bm{\mu}}^{\text{fw}}(\mathbf{Z}, \bm{\beta})$ a $\beta$-field estimator based on samples from $\mathbf{Z}$ and on (possibly non-optimal) field weights $\bm{\beta}$.
$\widehat{\bm{\mu}}^{\text{fw}}(\mathbf{Z}, \bm{\beta})$ is a possibly biased and possibly sub-optimal estimator for $\bm{\mu}_\alpha$.
Similarly, we denote by $\widehat{\bm{\mu}}^{\text{W}}(\mathbf{Z}, \bm{\beta}, \mathbf{W})$ the $W$-field estimator based on samples $\mathbf{Z}$ and using field weights $\bm{\beta}$.

Then equation \eqref{eq:def_Wfield} can be rewritten as
\begin{align}
\widehat{\bm{\mu}}^{\text{W}}(\mathbf{Z}, \bm{\beta}, \mathbf{W}) := \mathbf{W}\widehat{\bm{\mu}}^{\text{fw}}(\mathbf{Z}\mathbf{W}, \bm{\beta})\label{eq:Wfield-betafield}
\end{align}

\paragraph{Unbiasedness}
We first show that unbiased $W$-field estimators are necessarily associated to unbiased $\beta$-field estimator.
Let $(P1)$ be the proposition 
``$\widehat{\bm{\mu}}^{\text{W}}(\mat{Z}, \bm{\beta}, \mat{W})$ is an unbiased estimator of $\E[\mathbf{Z}]^\T\alpha$''.
\begin{align}
(P1)\;&\Longleftrightarrow&\E\left[\widehat{\bm{\mu}}^{\text{W}}(\mathbf{Z}, \bm{\beta}, \mathbf{W})\right] &= \E[\mathbf{Z}]^\T\alpha\\
&\Longleftrightarrow& \E\left[\mathbf{W}\widehat{\bm{\mu}}^{\text{fw}}(\mathbf{Z}\mathbf{W}, \bm{\beta})\right] &=\E[\mathbf{Z}]^\T\alpha\\
&\Longleftrightarrow& \mathbf{W}\E\left[\widehat{\bm{\mu}}^{\text{fw}}(\mathbf{Z}\mathbf{W}, \bm{\beta})\right] &=\E[\mathbf{Z}]^\T\alpha\\
&\Longleftrightarrow& \E\left[\widehat{\bm{\mu}}^{\text{fw}}(\mathbf{Z}\mathbf{W}, \bm{\beta})\right] &= \mathbf{W}^\T\E[\mathbf{Z}]^\T\alpha  \quad  \text{using }\mat{W}^\T\mat{W} = \mat{WW}^\T = \mat{I}_n\\
\quad&\Longleftrightarrow& \E\left[\widehat{\bm{\mu}}^{\text{fw}}(\mathbf{Z}\mathbf{W}, \bm{\beta})\right] &= \E[\mathbf{Z}\mathbf{W}]^\T\alpha\\
&\Longleftrightarrow& (P2)\qquad&
\end{align}
with $(P2)$: ``$\widehat{\bm{\mu}}^{\text{fw}}(\mathbf{Z}\mathbf{W}, \bm{\beta})$ is an unbiased estimator of $\E\bigl[\mathbf{Z}\mathbf{W}\bigr]^\T \alpha$''.

\paragraph{Minimal variance}
We then show that the mean square errors of two associated estimators are equal.
\begin{align}
\mse\bigl(\widehat{\bm{\mu}}^{\text{W}}(\mathbf{Z}, \bm{\beta}, \mathbf{W}), \E[\mathbf{Z}]^\T\alpha\bigr) &= \E\Bigl[\left\|\widehat{\bm{\mu}}^{\text{W}}(\mathbf{Z}, \bm{\beta}, \mathbf{W}) - \E[\mathbf{Z}]^\T\alpha\right\|^2\Bigr]\\
&= \E\Bigl[\left\|\mathbf{W}\widehat{\bm{\mu}}^{\text{fw}}(\mathbf{Z}\mathbf{W}, \bm{\beta}) - \E[\mathbf{Z}]^\T\alpha\right\|^2\Bigr]\\
&= \E\Bigl[\left\|\mat{W}^\T\mathbf{W}\widehat{\bm{\mu}}^{\text{fw}}(\mathbf{Z}\mathbf{W}, \bm{\beta}) -  \mat{W}^\T\E[\mathbf{Z}]^\T\alpha\right\|^2\Bigr]\label{eq:tmp74}\\
&= \E\Bigl[\left\|\widehat{\bm{\mu}}^{\text{fw}}(\mathbf{Z}\mathbf{W}, \bm{\beta}) - \E[\mathbf{Z}\mathbf{W}]^\T\alpha\right\|^2\Bigr]\label{eq:tmp75}\\
&= \mse\bigl( \widehat{\bm{\mu}}^{\text{fw}}(\mathbf{Z}\mathbf{W}, \bm{\beta}) , \E[\mathbf{Z}\mathbf{W}]^\T\alpha \bigr)
\end{align}
where we used $\mathbf{WW}^\T=\mathbf{I}_n$ to obtain \eqref{eq:tmp74} and $\mathbf{W}^\T\mathbf{W}=\mathbf{I}_n$ to obtain \eqref{eq:tmp75}.

Since the unbiased estimators have equal mean square errors, the best unbiased $W$-field estimator is associated to the best unbiased $\beta$-field estimator through \eqref{eq:Wfield-betafield}.
All results valid for the $\beta$-field estimator (optimal values, sample allocation) can thus be applied here, by replacing random vectors $\mathbf{Z}_{\ell}$ by $\mathbf{W}^\T\mathbf{Z}_{\ell}$.

\paragraph{Numerical cost of applying the estimator}
To limit the number of transforms, the $W$-field ML estimator can be applied as
\begin{align}
\widehat{\bm{\mu}}^{\text{W}}_\alpha = \mathbf{W}\sum_{k=1}^K\sum_{\ell\in S^{(k)}}\operatorname{Diag}\left(\bm{\beta}_\ell^{(k)}\right)\mathbf{W}^\T \Emc^{(k)}\bigl[\mathbf{Z}_\ell\bigr]
\end{align}
which counts $p$ forward transforms and one inverse transform (in the MLMC case, $2L$ transforms. See page \pageref{def:p} for the definition of $p$).

\paragraph{Numerical cost of estimating the optimal weights}
The estimation of the optimal weights requires the computation of element-wise covariance matrices $C^{(k,i)}$.
The simplest approach consists in estimating the $C^{(k,i)}$ from a set of $N\times L$ transformed and coupled realizations, with $N$ a large sampling size.
The transformation of this training ensemble into the $W$-space requires $N\times L$ forward transforms.

\paragraph{Choice of $\mathbf{W}$}
This estimator depends on the choice of $\mathbf{W}$.
With $\mathbf{W}=\mathbf{I}_n$, we retrieve the $\beta$-field estimator.
With a Fourier basis, we get the optimal spectral filters.

Note that if $\mathbf{W}$ is allowed to vary, the $W$-field estimators encompass all estimators with real valued symmetric matrices that are simultaneously diagonalizable.
This is less general than the BLUE introduced in section \ref{sec:nd-fullmatrixweights}, where the matrix weights are generally not symmetric.

\subsection{Matrix weights -- the multidimensional MLBLUE}
\label{sec:nd-fullmatrixweights}

The general MLBLUE for a random vector estimation has matrix weights, as mentioned in \citet{croci2023multifidelity}.
We follow the same approach as \SUml{} for the derivation of the MLBLUE in this context.
A variance minimization approach is also feasible and would yield the same results (see appendix \ref{app:nd_varminim}).

\paragraph{Notations}
Some notations need to be updated for this section:
\begin{itemize}
	\item The random vectors $\mathbf{Z}_1, \dots, \mathbf{Z}_L$ on each levels can now have different sizes $n_\ell$ possibly all different from $n$. 
	For instance, if the low-fidelity samples originate from simulations on coarse grids, there is no requirement to interpolate them back to the grid of the fine level $L$. 
	To shorten notations, we define:
	\begin{align}
		N &:= \sum_{\ell=1}^L n_\ell\\
		n^{(k)} &:= \sum_{\ell \in S{(k)}} n_\ell
	\end{align}

	\item $\mathbf{Z}$ is now the concatenation of the $L$ random vectors $\mathbf{Z}_\ell\in\mathbb{R}^{n_\ell}$.
	\begin{align}
		\mathbf{Z}&:= \Bigl(\mathbf{Z}_1^\T \cdots \mathbf{Z}_L^\T\Bigr)^\T \in \mathbb{R}^N\\
		\bm{\mu} &:= \E\bigl[\mathbf{Z}\bigr]
	\end{align}

	\item The selection and extension operators are extended accordingly, to select only the part $\mathbf{Z}^{(k)}$ of $\mathbf{Z}$ which is relevant to some coupling group $k$.
	\begin{align}
		&\mathbf{Z}^{(k)} := \Bigl(\cdots \mathbf{Z}_\ell^\T \cdots\Bigr)^\T_{\ell\in S^{(k)}} \in \mathbb{R}^ {n^{(k)}}\\
		&\mathbf{R}^{(k)}
		\text{ of size } n^{(k)}\times N
		\text{ such that } \mathbf{R}^{(k)}\mathbf{Z} = \mathbf{Z}^{(k)}\\
		&\mathbf{P}^{(k)} := \mathbf{R}^{(k)\T}\\
	\end{align}

	\item The $\alpha$ coefficients are extended to a matrix $\bm{\alpha}$ representing a linear map from $\mathbf{R}^N$ to $\mathbf{R}^n$.
	In the most common case, we are interested in the high-fidelity level and $n=n_L$. 
	In this case, $\bm{\alpha}$ is the selection operator for this level.
	\begin{align}
		\bm{\alpha} &:= \Bigl(\mathbf{0}_{n\times n_1} \dots 
							  \mathbf{0}_{n\times n_{L-1}} \mathbf{I}_n\Bigr)\\
	\end{align}
	The quantity of interest here is $\bm{\mu}_\alpha := \bm{\alpha\mu} \in \mathbb{R}^{n}$.

\end{itemize}

\paragraph{Normal equations}
All samples are concatenated in a big column vector of ``observations'' $\underline{\mathbf{Z}} := \bigl(\bigl(\mathbf{Z}^{(k),i}\bigr)_{i=1}^{m^{(k)}}\bigr)_{k=1}^K$, where $\mathbf{Z}^{(k),i}\in\mathbb{R}^{n^{(k)}}$ is the $i$-th (random) sample on coupling group $k$.
The size of $\underline{\mathbf{Z}}$ is $\sum_{k=1}^Km^{(k)}n^{(k)}$.

The ``observation operator'' relating $\bm{\mu}$ and $\underline{\mathbf{Z}}$ is the column block vector $\mathbf{H} :=\bigl(\bigl(\mathbf{R}^{(k)}\bigr)_{i=1}^{m^{(k)}}\bigr)_{k=1}^K$.
Then
\begin{align}
\underline{\mathbf{Z}} = \mathbf{H}\bm{\mu} + \bm{\epsilon}\\
\text{with}\quad \bm{\epsilon} := \underline{\mathbf{Z}} - \mathbf{H}\bm{\mu}
\end{align}

We have the following properties about the noise $\bm{\epsilon}$.
\begin{align}
\E[\bm{\epsilon}] &= 0\\
\C(\bm{\epsilon}, \bm{\epsilon}) &= \C\bigl(\underline{\mathbf{Z}},\,\underline{\mathbf{Z}}\bigr)\\
&= \operatorname{Diag}_{k=1}^K\Bigl(\operatorname{Diag}_{i=1}^{m^{(k)}}\bigl(\mathbf{C}^{(k)}\bigr)\Bigr)\\
\text{with}\quad \mathbf{C}^{(k)}  &:= \C\bigl(\mathbf{Z}^{(k)}, \mathbf{Z}^{(k)}\bigr)
\end{align}

\paragraph{Invertibility of the covariance matrices}
Thereafter, we assume that the matrices $\mathbf{C}^{(k)}$ are non-singular. 

This may not be the case in some cases, and some care should be taken in defining the low-fidelity samples.
For instance, the assumption is not valid if some low fidelity samples on level $\ell$ are coarse grid simulations linearly interpolated to a finer grid.
In this situation, the extrapolated elements in $\mathbf{Z}_\ell$ can be expressed as a linear combination of the other elements it was extrapolated from, so that $\mathbf{C}^{(k)}$ has a non zero kernel and is singular. 

To meet the non-singularity assumption in this case, the low fidelity simulations on the coarse grid should be used as is, without any interpolation. 
It is the role of the MLBLUE to compute the optimal linear transform that best maps samples from this low-fidelity level to the finest grid. 
The matrix weights here act not only as optimal multilevel weights, but also as interpolators and smoothers.

\paragraph{Associated generalized least-squares problem}
\begin{align}
\min\limits_{\bm{\mu}\in\mathbf{R}^{N}} \| \underline{\mathbf{Z}} - \mathbf{H}\bm{\mu} \|^2_{\C(\bm{\epsilon}, \bm{\epsilon})^{-1}}
\end{align}
which could be decomposed as a sum over the coupling group $k$, as a consequence of the block-diagonal structure of $\C(\bm{\epsilon}, \bm{\epsilon})^{-1}$.

\paragraph{BLUE for the expectation of a random vector}
The solution is given by
\begin{align}
\widehat{\bm{\mu}}^{\text{mat},\star} &= \left(\mathbf{H}^\T\C(\bm{\epsilon}, \bm{\epsilon})^{-1}\mathbf{H}\right)^{-1} \mathbf{H}^\T\C(\bm{\epsilon}, \bm{\epsilon})^{-1} \underline{\mathbf{Z}}\\
&= \bm{\phi}^{-1}\mathbf{y}\\
\text{with}\quad\bm{\phi} &:= \sum_{k=1}^K m^{(k)} \mathbf{P}^{(k)} \left(\mathbf{C}^{(k)}\right)^{-1}\mathbf{R}^{(k)}\\
\text{and}\quad\mathbf{y} &:= \sum_{k=1}^K m^{(k)} \mathbf{P}^{(k)}\left(\mathbf{C}^{(k)}\right)^{-1}\Emc^{(k)}\Bigl[\mathbf{Z}^{(k)}\Bigr]
\end{align}

\paragraph{Partial estimation}
The BLUE for $\bm{\mu}_\alpha$ is actually $\bm{\alpha}\widehat{\bm{\mu}}^{\text{mat},\star}$.

\paragraph{Matrix weights}
The previous equations can be expanded to evidence the multilevel structure of the estimator.
\begin{align}
\bm{\alpha}\widehat{\bm{\mu}}^{\text{mat},\star} &=  \sum_{k=1}^K \bm{\beta}^{(k)} \Emc^{(k)}\left[\mathbf{Z}^{(k)}\right]\\
&=  \sum_{k=1}^K \sum_{\ell\in S^{(k)}}\bm{\beta}_\ell^{(k)} \Emc^{(k)}\left[\mathbf{Z}_\ell\right]\\
\bm{\beta}^{(k)} &= m^{(k)}\bm{\alpha}\bm{\phi}^{-1}\mathbf{P}^{(k)}\left(\mathbf{C}^{(k)}\right)^{-1}\,\in\mathbb{R}^{n\times n^{(k)}}\label{eq:betak_matrix}\\
 &=\Bigl(\cdots\; \bm{\beta}_\ell^{(k)} \cdots\Bigr)_{\ell \in S^{(k)}}\label{eq:betalk_matrix}
\end{align}

\paragraph{Variance and sample allocation}
The approach of \citet{croci2023multifidelity} to solve the MOSAP via semdidefinite programming is not directly applicable to this case. 
Some work on extending it would be of interest. 

For a given selection of groups of levels, the variance of the ML matrix-weighted estimator with optimal matrix weights is given by
\begin{align}
\operatorname{Tr}\C\left(\widehat{\bm{\mu}}^{\text{mat},\star}, \widehat{\bm{\mu}}^{\text{mat},\star}\right) &= \operatorname{Tr} \left(\bm{\alpha}\bm{\phi}^{-1}(m)\bm{\alpha}^\T\right)\label{eq:opt_var_mw}
\end{align}
which can be used to find the optimal sample allocation $m$ for this specific choice of levels and coupling structure.

With suboptimal matrix weights $\bm{\beta}^{(k)}$, the variance is given by
\begin{align}
\operatorname{Tr}\C\left(\widehat{\bm{\mu}}^\text{mat}, \widehat{\bm{\mu}}^\text{mat}\right) &= \sum_{k=1}^K m^{(k)} \operatorname{Tr}\left(\bm{\beta}^{(k)} \mathbf{C}^{(k)} \left(\bm{\beta}^{(k)}\right)^\T\right)
\end{align}

\paragraph{Remark} In the case of a weighted MLMC, the no-bias condition implies that last matrix weight $\bm{\beta}_{L,K}$ is $\mathbf{I}_n$.

\paragraph{Computational cost}
This approach is untractable as is for large-di\-men\-sion systems.
Indeed, estimating the matrix weights requires:
\begin{itemize}
	\item Estimating the covariance matrix $\mathbf{C}:=\C(\mathbf{Z},\mathbf{Z})$ of size $N$ by $N$, which is a $\mathcal{O}\bigl(n^2\bigr)$ task. 
	The covariance matrices $\mathbf{C}^{(k)}$ can then be extracted as $\mathbf{C}^{(k)} = \mathbf{R}^{(k)}\mathbf{CP}^{(k)}$.
	Some of these $\mathbf{C}^{(k)}$ are invertible only if $\mathbf{C}$ is estimated with at least $\max_k n^{(k)} + 1$ samples, which may be very expensive.
	A possible workaround to the singularity of $\mathbf{C}$ could be using some regularization technique such as covariance localization or ridging.
	\item Inverting the $\bm{\phi}$ matrix, of size $N$, and each $\mathbf{C}^{(k)}$ matrix of size $n^{(k)}$, or solving linear systems of associated sizes if the weights are directly applied to some MC estimate.
\end{itemize}

\paragraph{Making it tractable in large dimensions}
There are various ways forward to simplify this multilevel estimator to make it tractable in large dimension. 
If the simulations are attached to a physical space with some notion of distance, the interpretation of the matrix weights as interpolators and smoothers suggest that the $\bm{\beta}_\ell^{(k)}$ matrices could be imposed as sparse. 
This means that the estimation of an element on the fine grid should only involve low-fidelity elements that are within some maximum distance.
The matrix weights could be further simplified by imposing  some invariance by translation, or have some periodicity based on the underlying grids.

Alternatively, the simplification could be done the other way round.  
The inter-level and inter-element covariance matrices $\mathbf{C}^{(k)}$ could be computed only on pair of points within a given distance. 
Beyond this distance, the covariances would be assumed to be zero. 
This could be an interesting avenue for future research.

\paragraph{Relation with other multilevel estimators}

The multilevel estimators introduced in sections \ref{sec:nd-scalar} to \ref{sec:nd-Wfield} can be considered as special cases of the general matrix-weight estimator.
\begin{itemize}
\item The estimator with scalar weights is restricting itself to $\bm{\beta}_\ell^{(k)}$ matrices of the form $\beta_\ell^{(k)}\mathbf{I}_n$.
\item The $\beta$-field estimator is restricting itself to $\bm{\beta}_\ell^{(k)}$ matrices of the form $\operatorname{Diag}\Bigl(\bm{\beta}_\ell^{(k)}\Bigr)$ (where $\bm{\beta}_\ell^{(k)}$ is now a vector).
\item The $W$-field estimator is restricting itself $\bm{\beta}_\ell^{(k)}$ matrices that are all diagonalizable in the basis $\mathbf{W}$.
\item The tractable approaches mentioned in previous paragraph would restrict to sparse matrix weights, or kernel-based matrices. 
\end{itemize}

\clearpage
\section{Estimation of a scalar covariance}
\label{sec:cov-scalar}
Let $X=(X_\ell)_{\ell=1}^L$ and $Y=(Y_\ell)_{\ell=1}^L$ be random vectors gathering the scalar random variables $X_1, \ldots, X_L$ and $Y_1, \ldots, Y_L$.
We group the covariances of $X_\ell$ and $Y_\ell$ for each $\ell$ in the vector $c:=(\C(X_\ell, Y_\ell))_{\ell=1}^L$.
We are interested in a linear combination of the elements of $c$, of the kind $\alpha^\T c$, with $\alpha\in\mathbb{R}^n\setminus\{0\}$.
In practice, $\alpha=(0 \dots 0\;1)\in\mathbb{R}^L$.

Given some some coupling structure $\left(S^{(k)}, m^{(k)}\right)_{k=1}^K$, we are looking for the best unbiased estimator of $\alpha^\T c$ of the form
\begin{align}
\Cmc_\alpha^\text{ML} = \sum_{k=1}^K\sum_{\ell\in S^{(k)}} \beta_\ell^{(k)} \Cmc^{(k)}(X_\ell,Y_\ell)\label{eq:general_form_ml_scalar_cov}
\end{align}
where $\Cmc^{(k)}(X_\ell,Y_\ell)$ is the sample covariance estimator of $X_\ell$ and $Y_\ell$ based on $m^{(k)}$ coupled samples.
\begin{align}
\Cmc^{(k)}(X_\ell,Y_\ell) := \frac{m^{(k)}}{m^{(k)}-1}\Emc^{(k)}\left(\left(X_\ell-\Emc^{(k)}(X_\ell)\right)\left(Y_\ell-\Emc^{(k)}(Y_\ell)\right)\right)
\end{align}
This is an unbiased estimator for $\C(X_\ell, Y_\ell)$.

\paragraph{A linear estimator?}
The multilevel MC estimators of the form \eqref{eq:general_form_ml_scalar_cov} depend quadratically on the samples, and as such are not truly linear estimators. 
A multilevel best \emph{linear} unbiased estimator for the covariance does not exist in general.
However, estimators \eqref{eq:general_form_ml_scalar_cov} are still linear in the MC estimators they combine, which is why we use the name MLBLUE for covariance estimators as well. 
The regression approach could be used here to derive the MLBLUE, but the MC estimators on each coupling groups should be used instead of the samples.

\paragraph{Relation to the expectation case}
This problem is very similar to the problem faced in section \ref{sec:scalar-expectation} for the estimation of the expectation.
The two important points needed to extend the results are the unbiasedness of the MC estimators involved, and an expression of their covariances. 
\begin{itemize}
	\item $\diag\left(\Cmc^{(k)}\left(X^{(k)}, Y^{(k)}\right)\right)$ is an unbiased estimator for $R^{(k)}c$, similar to the expectation case where $\Emc^{(k)}\left[Z^{(k)}\right]$ is an unbiased estimator of $R^{(k)}\mu$.
	\item However, the covariance of MC estimators for the covariances does not simplify as well as the MC estimators for the mean. 
\end{itemize}

Let $\mathbb{C}^{(k)}$, of size $p^{(k)}$ by $p^{(k)}$, denote the covariance of the MC estimator on group $k$ with $m^{(k)}$ samples.
It differs from the covariance $C^{(k)}$ of the random vectors $Z^{(k)}$.
For the estimation of the mean, we have
\begin{align}
	\mathbb{C}^{(k)}&=\C\left(\Emc^{(k)}\left[Z^{(k)}\right], \Emc^{(k)}\left[Z^{(k)}\right]\right)\\
	&=1/m^{(k)} C^{(k)}
\end{align}
For the estimation of the covariance, we have
\begin{align}
\mathbb{C}^{(k)} := \C\left(\diag\left(\Cmc^{(k)}\left(X^{(k)}, Y^{(k)}\right)\right), \diag\left(\Cmc^{(k)}\left(X^{(k)}, Y^{(k)}\right)\right)\right)\label{eq:covariance_of_mc_covariance}
\end{align}
which is expanded later in this section.

\paragraph{MLBLUE results expressed as a function of the $\mathbb{C}^{(k)}$}
The solution to the constrained minimization problem is given by
\begin{align}
\beta^{(k)} &= \left(\mathbb{C}^{(k)}\right)^{-1}R^{(k)} \phi^{-1}\alpha \label{eq:optbeta_generalC1}\\
\text{with } \phi &:= \sum_{k=1}^K P^{(k)} \bigl(\mathbb{C}^{(k)} \bigr)^{-1}R^{(k)}.
\end{align}

The variance of the MLBLUE is $\alpha^\T \phi^{-1}\alpha$.
Note that this paragraph is valid for the estimation of the expectation, for the estimation of the covariance, but also for the estimation of any scalar statistic which admits unbiased Monte Carlo estimators. 

\paragraph{Estimating the $\mathbb{C}^{(k)}$}
Covariances $\mathbb{C}^{(k)}$ can be expressed as a function of the first centered-moments of $R^{(k)}X$ and $R^{(k)}Y$, isolating the dependence on the sample size $m^{(k)}$.
Let's drop the $k$ for the sake of clarity, and focus on the covariance matrix $\mathbb{C}$ associated to $X$ and $Y$.
It can be shown (equation 9 of \BMa) that for $\ell, \ell' \in\{1,\dots,L\}$, element $\ell,\ell'$ of $\mathbb{C}$ is
\begin{multline}
\mathbb{C}_{\ell,\ell'} = \frac{\mathbb{M}^4\left[X_\ell,X_\ell',Y_\ell,Y_\ell'\right]}{m^{(k)}} - \frac{\C\left(X_\ell,Y_\ell\right)\C\left(X_{\ell'},Y_{\ell'}\right)}{m^{(k)}}\\ + \frac{\C\left(X_{\ell},Y_{\ell'}\right)\C\left(Y_{\ell},X_{\ell'}\right) + \C\left(X_{\ell},X_{\ell'}\right)\C\left(Y_{\ell},Y_{\ell'}\right)}{m^{(k)}(m^{(k)}-1)}\label{eq:cov_of_cov}
\end{multline}
with $\mathbb{M}^4\left[X_1, X_2, X_3, X_4\right]:=\E\left[\Pi_{i=1}^4\left(X_i - \E\left[X_i\right]\right)\right]$.

In the case of variance estimation, when $X=Y$, equation \eqref{eq:cov_of_cov} simplifies to
\begin{align}
\mathbb{C}_{\ell,\ell'} = \frac{\mathbb{M}^4\left[X_\ell,X_\ell'\right]}{m^{(k)}} - \frac{\V\left(X_\ell\right)\V\left(X_{\ell'}\right)}{m^{(k)}} + \frac{2\C\left(X_{\ell},X_{\ell'}\right)^2}{m^{(k)}(m^{(k)}-1)}\label{eq:cov_of_var}
\end{align}
with $\mathbb{M}^4\left[X_1, X_2\right]:=\mathbb{M}^4\left[X_1, X_1, X_2, X_2\right]$ (the definition of $\mathbb{M}^4$ depends on the number of parameters it is given).

\paragraph{MOSAP}
The semidefinite programming approach to the MOSAP does not extends naturally here, as the matrix $\phi(m)$ no longer depends linearly on the sample sizes $m$. 
Note this will be the case for all multilevel estimators of the covariance introduced in this note, including multilevel estimators of covariance matrices. 

This could be circumvented by minimizing an upper bound of the variance that linearly depends on the inverses of the samples sizes, as done ny \citet{mycek2019multilevel}.

Another solution that does not introduce approximations is given by falling back to the constrained minimization of the non linear variance $\alpha^\T \phi(m)^{-1}\alpha$, or by solving the non-linear semidefinite problem of \citet{croci2023multifidelity}.

\paragraph{Computational cost}
\begin{itemize}
\item In practice, all the involved matrices are of size at most $L$ (\textit{i.e.} 2 or 3).
\item Ensemble estimates of fourth-order moments are more noisy than ensemble covariances.
As a consequence, the ensemble sizes used to estimate $\mathbb{C}$ should be larger here than in section \ref{sec:scalar-expectation} to reach a similar robustness.
\item $\mathbb{C}$ is a symmetric matrix, so only $L(L+1)/2$ elements really need to be estimated. This number can be even more reduced by noting that covariances between levels that are not coupled ($\ell, \ell'$ so that $\forall k,\,\ell\in S^{(k)} \Rightarrow \ell'\notin S^{(k)}$) need not be computed.
\end{itemize}

\clearpage
\section{Estimation of a covariance matrix}
\label{sec:cov-nd}

\subsection{The problem}
How do the results of previous section extend to the multidimensional case?
The general MLBLUE for the estimation of a covariance matrix involves fourth-order tensors linearly combining the entries of several covariance matrix estimators into one matrix.
We don't derive it here, as it is computationnally prohibitively expensive in high-dimension applications. 
Instead, we focus on multilevel estimators of the covariance matrix that are both simpler and computationally affordable.

\paragraph{}
The $\mathbf{X}_\ell$ are now random vectors of $\Omega\to\mathbb{R}^n$.
Note that they are imposed to have the same dimension, which would not be the case with the general MLBLUE.
They are concatenated in the random vector $\mathbf{X}:= (\mathbf{X}_1^\T,\dots,\mathbf{X}_L^\T)^\T$.
We are interested in the covariance matrix $\C(\mathbf{X}, \mathbf{X})$ of size $nL$.

\paragraph{Partial estimation}
More specifically, we are often interested in estimating only some blocks of $\C(\mathbf{X}, \mathbf{X})$, for instance the last block $\C(\mathbf{X}_L, \mathbf{X}_L)$.
To extend the formalism to this kind of situation, we focus on estimating $\mathbf{C}^\alpha:=\sum_{\ell=1}^L\alpha_\ell\C(\mathbf{X}_\ell, \mathbf{X}_\ell)$ for some non-zero vector $\alpha\in\mathbb{R}^L$.

\paragraph{Generalized variance for a matrix estimator}
We choose the Frobenius norm to measure mean square errors for matrix estimators, as in \BMa.
The choice of the Frobenius norm is associated to the following bias-variance decomposition:
\begin{align}
\mse\left(\widehat{\mathbf{B}}, \mathbf{B}\right) = \sum_{i=1}^n\sum_{j=1}^n \V(\widehat{B}_{ij}) + \left\|\E\left[\widehat{\mathbf{B}} - \mathbf{B}\right]\right\|^2_\mathrm{F}
\end{align}

Hereafter, we overload the variance operator and define
\begin{align}
	\V(\widehat{\mathbf{B}}):= \sum_{i=1}^n\sum_{j=1}^n \V(\widehat{B}_{ij})
\end{align}
for any matrix-valued random variable $\widehat{\mathbf{B}}$.

\subsection{Scalar weights}
\label{sec:cov-nd-scalar}
We first consider the case of scalar weights, one for each MC estimator, common to all matrix elements in the estimator.
\paragraph{Class of estimators}
\begin{align}
\bCmc^\text{ML} = \sum_{k=1}^K \sum_{\ell\in S^{(k)}} \beta_\ell^{(k)} \Cmc^{(k)}\left(\mathbf{X}_\ell, \mathbf{X}_\ell\right)\label{eq:sc_covmat}
\end{align}
where $\beta_\ell^{(k)}$ is a scalar.

\paragraph{Optimal scalar weights}
The optimal weights can be retrieved very similarly to what is done for multidimensional expectation in section \ref{sec:nd-scalar}.

The same relation \eqref{eq:nobias_sc} guarantees the unbiasedness of the estimator.
It can be used as a constraint to minimize the variance of the estimator, given by a relationship similar to equation \eqref{eq:var_sc}.
\begin{align}
	\V\left(\bCmc^\text{ML}\right) &= \sum_{k=1}^K\left(\beta^{(k)}\right)^\T \left(\sum_{i=1}^n\sum_{j=1}^n \mathbb{C}^{(k, ij)}\right)\beta^{(k)}\label{eq:var_sc_covmat}
\end{align}
where $\mathbb{C}^{(k, ij)}$ is the covariance matrix of size $p^{(k)}\times p^{(k)}$ for Monte Carlo covariance estimators $\widehat{C}^{(k)}(X_{\ell,i}, X_{\ell,j}),\,\ell\in S^{(k)}$.

The optimal weights are then given by equation \eqref{eq:optbeta_generalC1}, replacing $\mathbb{C}^{(k)}$ with\break $\sum_{i=1}^n\sum_{j=1}^n \mathbb{C}^{(k, ij)}$, or equivalently with $1/n^2\sum_{i=1}^n\sum_{j=1}^n \mathbb{C}^{(k, ij)}$.

\paragraph{Estimating the average $\mathbb{C}^{(k, ij)}$}
The expression of the average inter-level covariance matrix of MC estimators directly follows from the expression in the scalar case (section \ref{sec:cov-scalar}).
\begin{multline}
\sum_{i=1}^n\sum_{j=1}^n \mathbb{C}^{(k, ij)}_{\ell,\ell'} = \sum_{i=1}^n\sum_{j=1}^n\left(\frac{\mathbb{M}^4\left[X_{\ell,i},X_{\ell',i},X_{\ell,j},X_{\ell',j}\right]}{m^{(k)}}\right.\\
- \frac{\C\left(X_{\ell,i},X_{\ell,j}\right)\C\left(X_{\ell',i},X_{\ell',j}\right)}{m^{(k)}} \\
\left. + \frac{\C\left(X_{\ell,i},X_{\ell',j}\right)\C\left(X_{\ell,j},X_{\ell',i}\right) + \C\left(X_{\ell,i},X_{\ell',i}\right)\C\left(X_{\ell,j},X_{\ell',j}\right)}{m^{(k)}{(m^{(k)}-1)}}\right)\label{eq:averageCC}
\end{multline}

\paragraph{MOSAP}
As explained in section \ref{sec:cov-scalar}, the semidefinite programming approach to the MOSAP does not extend naturally here. 

\paragraph{Computational cost}
Estimating $\sum_{i=1}^n\sum_{j=1}^n \mathbb{C}^{(k, ij)}$ directly is not tractable in high dimension, as the number of operations is quadratic in the grid size $n$. 
Fortunately, it is possible to reduce the cost of this estimation from $\mathcal{O}\left(n_en^2\right)$ to $\mathcal{O}\left(n_e^2n\right)$, where $n_e$ is the size of the ensemble used to estimate centered statistics in the pre-processing step (see appendix \ref{app:covcov_estimate}).

\paragraph{Sample allocation for MLMC}
The expression of the variance of a general estimator given here can be used to estimate the optimal sample allocation for the MLMC estimator of a covariance matrix.
This is detailed in appendix \ref{app:opt-alloc-mlmc-cov-mat}.

\subsection{Matrix field weights}
\label{sec:cov-nd-matrixfield}
\paragraph{Class of estimators}
\begin{align}
\bCmc^\text{ML} = \sum_{k=1}^K \sum_{\ell\in S^{(k)}} \bm{\beta}_\ell^{(k)} \circ \Cmc^{(k)}\left(\mathbf{X}_\ell, \mathbf{X}_\ell\right) \label{eq:covmat-gal-class}
\end{align}
where $\bm{\beta}_\ell^{(k)}$ is an $n \times n$ matrix and $\circ$ denotes the Schur (element-wise) product.

\paragraph{Unbiasedness, Variance, Optimal weights}
From the definition of the generalized variance derived from the Frobenius norm, the problem can be decomposed as independent estimations of $n^2$  scalar covariances.
See section \ref{sec:cov-scalar} for more details.

\paragraph{Numerical cost}
Unaffordable as such, since it requires the estimation of $n^2$ scalar coefficients.

\paragraph{Relation to covariance localization}
The use of a Schur product applied to a covariance matrix in equation \eqref{eq:covmat-gal-class} reminds of covariance localization in data assimilation \citep[e.g.][]{lorenc2003potential}.
This regularization technique consists in element-wise multiplication of an ensemble covariance matrix with a parametrized distance-dependent correlation matrix $\mathbf{L}$:
\begin{align}
	\mathbf{L}\circ \Cmc^{(k)}\left(\mathbf{X}_\ell, \mathbf{X}_\ell\right)
\end{align}
Localization differs from what has been considered so far, has it yields a \emph{biased} estimator of the covariance matrix wherever $\mathbf{L}$ is not 1. 
Optimizing the variance without the bias constraint makes no sense, but optimizing the MSE does, since it includes a bias and a variance contribution.
This is one of the ideas presented in \BMb, which we extend to the multilevel case in next section.

\subsection{Optimal localization}
\label{sec:opt-loc}
We propose here an extension of the optimal localization theory of \BMa{} and \BMb{} to multilevel covariance estimators.
We first derive the results in the line of the previous sections.
An aptempt to propose a derivation based on the approach of \BMa{} and \BMb{} (see also \citealp{menetrier2020covariance}) is proposed in appendix \ref{app:opt-loc}.
Though both approaches yield the same practical results, the associated assumptions and interpretations are slightly different.

\paragraph{}
Contrarily to previous sections, we no longer minimize the variance under a no-bias constraint.
The localization makes the covariance estimator biased, so we rather minimize the mean squared error of the localized multilevel estimator.

\subsubsection{General case}
\label{sec:optloc_general}
To simplify the notations, we note $\mathbf{B}_\ell:=\C(\mathbf{X}_\ell, \mathbf{X}_\ell)$ and $\widetilde{\mathbf{B}}_\ell^{(k)}:=\Cmc^{(k)}\left(\mathbf{X}_\ell, \mathbf{X}_\ell\right)$.
From the unbiasedness of the Monte Carlo covariance estimator, we have $\mathbb{E}\left[\widetilde{\mathbf{B}}_\ell^{(k)}\right] = \mathbf{B}_\ell$.

\paragraph{Class of estimators}
\begin{align}
\widehat{\mathbf{B}}^\text{ML} = \sum_{k=1}^K \sum_{\ell\in S^{(k)}} \mathbf{L}_\ell^{(k)} \circ \widetilde{\mathbf{B}}_\ell^{(k)}
\end{align}
where $\mathbf{L}_\ell^{(k)}$ is an $n \times n$ matrix, without any imposed structure at this stage.

\paragraph{Minimizing the MSE}
We want to minimize the mean square error of the localized covariance estimator:
\begin{align}
\mse\left(\widehat{\mathbf{B}}^\text{ML}, \mathbf{B}_L\right) &= \sum_{i=1}^n\sum_{j=1}^n\mathbb{E}\left[\left(\sum_{k=1}^K \sum_{\ell\in S^{(k)}} {L}_{\ell,ij}^{(k)} \circ \widetilde{{B}}_{\ell,ij}^{(k)} - {B}_{L,ij}\right)^2\right]\label{eq:fullmse_optloc}
\end{align}
The function to minimize is the sum of $n^2$ independent cost functions.
We now focus on one independent sub-problem $i,j$.
To ease the notations, we drop the $ij$ indexes and denote $L_\ell^{(k)}:=L_{\ell,ij}^{(k)}$, $\widetilde{B}_\ell^{(k)} := \widetilde{B}_{\ell,ij}^{(k)}$ and $B_{\ell} := B_{\ell,ij}$.
The problem reads as
\begin{align}
\underset{L_\ell^{(k)},\, 1\leq k\leq K,\,\ell\in S^{(k)}}{\min} \;\mathbb{E}\left[\left(\sum_{k=1}^K \sum_{\ell\in S^{(k)}} {L}_\ell^{(k)} \widetilde{{B}}_{\ell}^{(k)} - {B}_{L}\right)^2\right]\label{eq:optloc-subproblem}
\end{align}
To further simplify the notations, we define stacked vectors containing the information from all levels in all coupling groups:
\begin{align}
\underline{L}&:=\left(\left(L_\ell^{(k)}\right)_{\ell\in S^{(k)}}\right)_{k=1}^K\in \mathbb{R}^p\\
\underline{\widetilde{B}}&:=\left(\left(\widetilde{B}_\ell^{(k)}\right)_{\ell\in S^{(k)}}\right)_{k=1}^K\in \mathbb{R}^p\\
\underline{B}&:=\left(\left(B_\ell\right)_{\ell\in S^{(k)}}\right)_{k=1}^K\in \mathbb{R}^p
\end{align}
with $p=\sum_{k=1}^K p^{(k)}$.

Equation \eqref{eq:optloc-subproblem} becomes
\begin{align}
\underset{\underline{L}}{\min}\;\mathbb{E}\left[\left(\underline{L}^\T \underline{\widetilde{{B}}} - {B}_{L}\right)^2\right] &= \underset{\underline{L}}{\min}\;\E\left[ \left(\underline{L}^\T\underline{\widetilde{{B}}}\right)\left(\underline{\widetilde{{B}}}^\T\underline{L}\right) - 2B_{L}\left(\underline{\widetilde{{B}}}^\T\underline{L}\right) + B_{L}^2\right]\nonumber\\
&= \underset{\underline{L}}{\min}\;\underline{L}^\T\E\left[ \underline{\widetilde{{B}}}\underline{\widetilde{{B}}}^\T\right]\underline{L} - 2B_{L}\E\left[\underline{\widetilde{{B}}}^\T\right]\underline{L} + B_{L}^2\nonumber\\
&= \underset{\underline{L}}{\min}\;\underline{L}^\T\E\left[ \underline{\widetilde{{B}}}\underline{\widetilde{{B}}}^\T\right]\underline{L} - 2B_{L}\underline{B}^\T\underline{L}
\end{align}
Setting the gradient with respect to $\underline{L}$ to zero yields the optimality criterion
\begin{align}
\E\left[ \underline{\widetilde{{B}}}\underline{\widetilde{{B}}}^\T\right]\underline{L} = B_{L}\underline{B}
\end{align}

\paragraph{Invertibility of $\E\left[ \underline{\widetilde{{B}}}\underline{\widetilde{{B}}}^\T\right]$}
The uniqueness of the optimal localization is not guaranteed, in particular if the matrix $\E\left[ \underline{\widetilde{{B}}}\underline{\widetilde{{B}}}^\T\right]$ on the left-hand side is not invertible.
This should not be a problem though, as cases of non-invertibility are related to flat gradient in the MSE, which means to variations of $\underline{L}$ which don't impact the MSE. 
In practice, the non-uniqueness of these cases $i,j$ can be decided from the neighbouring pairs of points. 
For instance, parametric correlation matrices can be fit to these raw optimal localizations, to ensure the positive semidefiniteness of the localization matrices. 

\paragraph{Sample or asymptotic quantities}
The left hand-side matrix could be expressed as a function of the sample size and asymptotic quantities.
We could then use a large independent ensemble to estimate these asymptotic quantities.
In practice, we don't have such large ensembles.
A possible workaround is to express everything in terms of expectations of sample quantities, and to exploit the structure of the localization matrix.

\subsubsection{Imposing some structure to the localization matrix}
In practice, the localization matrix has some predefined structure, both for ease of estimation and ease of use.
The state space $\mathbb{R}^n$ is attached to some physical space, three-dimensional for instance.
Assuming for instance horizontal homogeneity and isotropy, we can define an equivalence relation between pairs $(i,j)$, associating pairs that should have the same localization.
The space of pairs $\mathbb{R}^n\times\mathbb{R}^n$ is thus partitioned into equivalence classes.
\begin{align}
L_{\ell,ij}^{(k)} = L_{\ell,\mathcal{C}}^{(k)}\text{ where }\mathcal{C} = \left[(i,j)\right]\text{ is the equivalence class of }(i,j)
\end{align}

In this context, equation \eqref{eq:fullmse_optloc} reads as
\begin{align}
\mse\left(\widehat{\mathbf{B}}^\text{ML}, \mathbf{B}_L\right) &= \sum_{i=1}^n\sum_{j=1}^n\mathbb{E}\left[\left(\sum_{k=1}^K \sum_{\ell\in S^{(k)}} {L}_{\ell,[(i,j)]}^{(k)} \circ \widetilde{{B}}_{\ell,ij}^{(k)} - {B}_{L,ij}\right)^2\right]
\end{align}
where the localization ${L}_{\ell,ij}^{(k)}$ has been replaced by ${L}_{\ell,[(i,j)]}^{(k)}$.
This minimization problem is associated to independent problems for each class $\mathcal{C}=[(i,j)]$.
\begin{align}
\underset{L_{\ell,\,\mathcal{C}}^{(k)},\, 1\leq k\leq K,\,\ell\in S^{(k)}}{\min} \;\sum_{(i,j)\in\mathcal{C}}\mathbb{E}\left[\left(\sum_{k=1}^K \sum_{\ell\in S^{(k)}} {L}_{\ell,\,\mathcal{C}}^{(k)} \widetilde{{B}}_{\ell,ij}^{(k)} - {B}_{L,ij}\right)^2\right]
\end{align}
As previously done, we define stacked vectors of $\mathbb{R}^p$: $\underline{L}$, $\underline{\widetilde{B}_{ij}}$ and $\underline{B_{ij}}$ for $(i,j)\in\mathcal{C}$.
Expanding the MSE gives
\begin{multline}
\underset{\underline{L}}{\min}\;\sum_{(i,j)\in\mathcal{C}}\mathbb{E}\left[\left(\underline{L}^\T \underline{\widetilde{B}_{ij}} - {B}_{L,ij}\right)^2\right] \\
= \underset{\underline{L}}{\min}\;\underline{L}^\T\left(\sum_{(i,j)\in\mathcal{C}}\E\left[ \underline{\widetilde{B}_{ij}}\underline{\widetilde{B}_{ij}}^\T\right]\right)\underline{L} - 2\left(\sum_{(i,j)\in\mathcal{C}}B_{L,ij}\underline{B_{ij}}^\T\right)\underline{L}
\end{multline}
which is associated to the optimality criterion
\begin{align}
\left(\frac{1}{\left|\mathcal{C}\right|}\sum_{(i,j)\in\mathcal{C}}\E\left[ \underline{\widetilde{B}_{ij}}\underline{\widetilde{B}_{ij}}^\T\right]\right)\underline{L} = \left(\frac{1}{\left|\mathcal{C}\right|}\sum_{(i,j)\in\mathcal{C}}B_{L,ij}\underline{B_{ij}}^\T\right)\label{eq:optimality_classes}
\end{align}

\paragraph{Relating asymptotic quantities to expectations of sampled moments}
Using results from \citet[equation 4.2]{menetrier2020covariance}, we can relate asymptotic quantities to expectations of sample quantities:
\begin{multline}
B_{L,ij}B_{\ell,ij} = P_1(m)\E\left[\Cmc_m\left({X}_{\ell,i}, {X}_{\ell,j}\right)\Cmc_m\left({X}_{L,i}, {X}_{L,j}\right)\right] \\
+ P_2(m)\left(\E\left[\Cmc_m\left({X}_{\ell,i}, {X}_{L,i}\right)\Cmc_m\left({X}_{\ell,j}, {X}_{L,j}\right)\right]\right.\\
+\left.\E\left[\Cmc_m\left({X}_{\ell,i}, {X}_{L,j}\right)\Cmc_m\left({X}_{\ell,j}, {X}_{L,i}\right)\right]\right)\\
+ P_3(m)\E\left[\widehat{M}^{4}_m\left({X}_{\ell,i}, {X}_{\ell,j}, {X}_{L,i}, {X}_{L,j}\right)\right]\label{eq:asy2sample}
\end{multline}
where $m\geq 4$ is a sampling size, $\Cmc_m$ is the unbiased Monte Carlo estimator of covariance for $m$ samples,  $\widehat{M}^{4}_m$ is the Monte Carlo estimator for fourth-order centered moments, and $P_1(m)$, $P_2(m)$ and $P_3(m)$ are rational fractions.
\begin{align}
\Cmc_m(X, Y) &= \frac{1}{m-1}\sum_{i=1}^m\widetilde{X}^{(i)}\widetilde{Y}^{(i)}\\
\widehat{M}^{4}_m(X, Y, Z, T) &= \frac{1}{m}\sum_{i=1}^m\widetilde{X}^{(i)}\widetilde{Y}^{(i)}\widetilde{Z}^{(i)}\widetilde{T}^{(i)}\\
\text{with}\quad \widetilde{X}^{(i)} &= X^{(i)} - \widehat{E}_m(X) \quad \text{etc.}\\
P_1(m) &= \frac{(m-1)(m^2-3m+1)}{m(m-2)(m-3)}\\
P_2(m) &= \frac{m-1}{m(m-2)(m-3)}\\
P_3(m) &= -\frac{m}{(m-2)(m-3)}
\end{align}

In practice, this relation requires estimating covariances between fine level $L$ and any level $\ell$.
This can be done using a coupling group involving all fidelity levels.
Ideally, the estimation should be performed independently of the covariance estimation.

\paragraph{Ergodicity assumption}
Using expression \eqref{eq:asy2sample} to express the right-hand side of \eqref{eq:optimality_classes}, we still have to express expectations of sample quantities.
Estimating this expectations by single-sample Monte Carlo estimators is a possible solution.
The high resulting sampling noise is actually cancelled out by the averaging over the whole equivalence class $\mathcal{C}$.
Pretending that the noise is averaged out implicitly supposes that space averages are equivalent to sampling a common process (ergodic assumption).
The approach of Benjamin Ménétrier presented in appendix \ref{app:opt-loc} gives more insight on what this process may be.

\paragraph{Numerical cost}
In practice, the operator $\sum_{(i,j)\in\mathcal{C}}\E\left[\cdot\right]$ could also be replaced by an average over a random subset of $\mathcal{C}$ as is done in \BMa.

\paragraph{Drawbacks}
In a single-level setting, \citet{menetrier2015optimized} showed how the localization and hybridization weights could be jointly optimized.
The approach was very appealing, but later trials showed it was less robust than a two-step optimization, where localization would be first optimized before optimizing the hybridization weights (M\'{e}n\'{e}trier, personal communication).
A similar behavior may occur here, where having too many degrees of freedom in the optimization problem may result in non robust solutions.

\clearpage
\appendix
\section{Retrieving the MLBLUE for the multidimensional expectation through constrained minimization of the variance}
\label{app:nd_varminim}
We retrieve here the results of section \ref{sec:nd-fullmatrixweights} with the variance minimization approach used in \ref{sec:scalar-expectation}.
The notations of section \ref{sec:nd-fullmatrixweights} are used here.
In particular, all random quantities are column vectors.

We define the class of matrix-weighted ML estimators for $\bm{\alpha\mu}=\bm{\mu}_\alpha$ of the form
\begin{align}
\widehat{\bm{\mu}}^{\text{mat}} = \sum_{k=1}^K\sum_{\ell\in S^{(k)}} \bm{\beta}_{\ell}^{(k)} \Emc^{(k)}\bigl[\mathbf{Z}_\ell\bigr]
\end{align}
where the $\bm{\beta}_{\ell}^{(k)}$ are matrices of $\mathbb{R}^{n\times n_\ell}$.
The inner sum can be made implicit by joining these matrices into $\bm{\beta}^{(k)} := \bigl(\cdots{}\; \bm{\beta}_{\ell}^{(k)} \cdots{}\bigr)_{\ell\in S^{(k)}}$.
\begin{align}
\widehat{\bm{\mu}}^{\text{mat}} = \sum_{k=1}^K\bm{\beta}^{(k)} \Emc^{(k)}\bigl[\mathbf{Z}^{(k)}\bigr]
\end{align}

\paragraph{Variance of the estimator}
The variance of the matrix-weighted estimator is given by
\begin{align}
\operatorname{Tr}\C\bigl(\widehat{\bm{\mu}}^{\text{mat}}, \widehat{\bm{\mu}}^{\text{mat}}\bigr) &= \sum_{k=1}^K \frac{1}{m^{(k)}}\bm{\beta}^{(k)} \mathbf{C}^{(k)} \left(\bm{\beta^{(k)}}\right)^\T \\
& = \bm{\beta} \operatorname{Diag}_{k=1}^K\biggl(\frac{1}{m^{(k)}}\mathbf{C}^{(k)}\biggr)\bm{\beta}^\T\\
\text{with}\quad \mathbf{C}^{(k)}  &:= \C\bigl(\mathbf{Z}^{(k)}, \mathbf{Z}^{(k)}\bigr)\\
\text{and}\quad \bm{\beta} &:= \bigl(\cdots\; \bm{\beta}^{(k)}\cdots\bigr)_{1\leq k\leq K}\in\mathbf{R}^{n\times \sum_k n^{(k)}}
\end{align}

\paragraph{Unbiasedness constraint}
The no-bias condition is given by
\begin{align}
\sum_{k=1}^K \bm{\beta}^{(k)} \mathbf{R}^{(k)} - \bm{\alpha} &= 0\\
\Leftrightarrow\qquad \bm{\beta}\mathbf{P}^\T - \bm{\alpha} &= 0\\
\text{with}\quad \mathbf{P} &:= \left(\mathbf{P}^{(1)} \dots \mathbf{P}^{(K)}\right)\in\mathbf{R}^{N\times \sum_k n^{(k)}}
\end{align}

\paragraph{Unconstrained minimization problem}
We introduce Lagrange multipliers $\Lambda$.
\begin{multline}
\bm{\beta}, \bm{\Lambda} = \underset{\bm{\beta}\in\mathbf{R}^{p\times \sum_k n^{(k)}}, \bm{\Lambda}\in \mathbb{R}^{n\times N}}{\operatorname{argmin}} \frac{1}{2}\operatorname{Tr}\biggl(\bm{\beta} \operatorname{Diag}_{k=1}^K\biggl(\frac{1}{m^{(k)}}\mathbf{C}^{(k)}\biggr)\bm{\beta}^\T\biggr)\\ - \sum_{i=1}^n\sum_{j=1}^{N} \Lambda_{ij} \left(\bm{\beta}\mathbf{P}^\T - \bm{\alpha}\right)_{ij}
\end{multline}

\paragraph{Associated linear system}
Setting the gradient with respect to $\bm{\beta}$ and $\bm{\Lambda}$ to zero yields the following linear system.
\begin{align}
\bm{\beta}\bm{\Sigma} - \bm{\Lambda}\mathbf{P} &= \mathbf{0}\\
\bm{\beta}\mathbf{P}^\T &= \bm{\alpha}
\end{align}
with $\bm{\Sigma} := \operatorname{Diag}_{k=1}^K\bigl(\frac{1}{m^{(k)}}\mathbf{C}^{(k)}\bigr)$.

\paragraph{Optimal matrix weights}
The unique solution is given by
\begin{align}
\bm{\beta} &= \bm{\alpha}\bm{\phi}^{-1}\mathbf{P}\operatorname{Diag}_{k=1}^K\Bigl(m^{(k)}\left(\mathbf{C}^{(k)}\right)^{-1}\Bigr)\\[2ex]
\text{with}\quad\bm{\phi} &:= \mathbf{P}\operatorname{Diag}_{k=1}^K\Bigl(m^{(k)}\bigl(\mathbf{C}^{(k)}\bigr)^{-1}\Bigr)\mathbf{P}^\T \\
&= \sum_{k=1}^K m^{(k)}\mathbf{P}^{(k)} \left(\mathbf{C}^{(k)}\right)^{-1}\mathbf{R}^{(k)}
\end{align}
From which we can retrieve expressions \eqref{eq:betak_matrix} and \eqref{eq:betalk_matrix} for $\bm{\beta}^{(k)}$ and $\bm{\beta}_{\ell}^{(k)}$.

\clearpage
\section{Estimating the average covariance matrix of covariance estimators}
\label{app:covcov_estimate}

The elements of the average covariance matrix $\sum_{i=1}^n\sum_{j=1}^n \mathbb{C}^{(k, ij)}$ of covariance estimators, given by equation \eqref{eq:averageCC}, are needed in two (not unrelated) situations:
\begin{enumerate}
\item To numerically optimize the weights and generalized sample allocation of the MLBLUE of a covariance matrix with scalar weights (see \ref{sec:cov-nd-scalar});
\item To numerically optimize the generalized sample allocation of an MLMC estimator of a covariance matrix (see appendix \ref{app:opt-alloc-mlmc-cov-mat}).
\end{enumerate}
The goal of this appendix is to provide computationally tractable ways to estimate the $L\times L$ matrix of these elements.

\paragraph{}
Each element of the matrix of interest can be decomposed into four terms.
For $1\leq k\leq K$ and $\ell, \ell'\in S^{(k)}$:
\begin{multline}
\sum_{i=1}^n\sum_{j=1}^n \mathbb{C}^{(k, ij)}_{\ell,\ell'} = \sum_{i=1}^n\sum_{j=1}^n\frac{\mathbb{M}^4\left[X_{\ell,i},X_{\ell',i},X_{\ell,j},X_{\ell',j}\right]}{m^{(k)}}\\
- \sum_{i=1}^n\sum_{j=1}^n\frac{\C\left(X_{\ell,i},X_{\ell,j}\right)\C\left(X_{\ell',i},X_{\ell',j}\right)}{m^{(k)}} \\
+ \sum_{i=1}^n\sum_{j=1}^n\frac{\C\left(X_{\ell,i},X_{\ell',j}\right)\C\left(X_{\ell,j},X_{\ell',i}\right)}{m^{(k)}{(m^{(k)}-1)}}\\
+ \sum_{i=1}^n\sum_{j=1}^n\frac{\C\left(X_{\ell,i},X_{\ell',i}\right)\C\left(X_{\ell,j},X_{\ell',j}\right)}{m^{(k)}{(m^{(k)}-1)}}
\end{multline}
As is, the double sums over the state dimension $n$ prevent any explicit computation.
Fortunately, naive Monte Carlo estimates of the centred moments can be reordered to make sure the cost of estimation stays linear in $n$.

\paragraph{}
Each one of the four terms can be estimated from (biased) Monte Carlo estimates based on coupled independent samples indexed by $s$, $1\leq s\leq n_e$.
We denote as $\widetilde{X}^{s}_{\ell',i}:=X_{\ell,i}^{s} - \frac{1}{n_e}\sum_{s=1}^{n_e}X_{\ell,i}^{s}$ the centred perturbations associated to a sample $X_{\ell,i}^{s}$.

\paragraph{First term: fourth-order moment}
\begin{align}
\sum_{i=1}^n\sum_{j=1}^n\mathbb{M}^4\left[X_{\ell,i},X_{\ell',i},X_{\ell,j},X_{\ell',j}\right] &\approx  \sum_{i=1}^n\sum_{j=1}^n\frac{1}{n_e}\sum_{s=1}^{n_e} \widetilde{X}^{s}_{\ell,i}\widetilde{X}^{s}_{\ell',i}\widetilde{X}^{s}_{\ell,j}\widetilde{X}^{s}_{\ell',j}\\
&=\frac{1}{n_e}\sum_{s=1}^{n_e}\left(\sum_{i=1}^n\widetilde{X}^{s}_{\ell,i}\widetilde{X}^{s}_{\ell',i}\right)^2
\end{align}

\paragraph{Second term: product of intra-level covariances}
\begin{multline}
\sum_{i=1}^n\sum_{j=1}^n \C\left(X_{\ell,i},X_{\ell,j}\right)\C\left(X_{\ell',i},X_{\ell',j}\right)\approx\\
\sum_{i=1}^n\sum_{j=1}^n \left(\frac{1}{n_e-1}\sum_{s=1}^{n_e}\widetilde{X}^{s}_{\ell,i}\widetilde{X}^{s}_{\ell,j}\right)\left(\frac{1}{n_e-1}\sum_{s'=1}^{n_e}\widetilde{X}^{s'}_{\ell',i}\widetilde{X}^{s'}_{\ell',j}\right)\\
= \frac{1}{(n_e-1)^2}\sum_{s=1}^{n_e}\sum_{s'=1}^{n_e}\left(\sum_{i=1}^n\widetilde{X}^{s}_{\ell,i}\widetilde{X}^{s'}_{\ell',i}\right)^2
\end{multline}

\paragraph{Third term: product of inter-level inter-point covariances}
\begin{multline}
\sum_{i=1}^n\sum_{j=1}^n\C\left(X_{\ell,i},X_{\ell',j}\right)\C\left(X_{\ell,j},X_{\ell',i}\right)\approx \\
\sum_{i=1}^n\sum_{j=1}^n \left(\frac{1}{n_e-1}\sum_{s=1}^{n_e}\widetilde{X}^{s}_{\ell,i}\widetilde{X}^{s}_{\ell',j}\right)\left(\frac{1}{n_e-1}\sum_{s'=1}^{n_e}\widetilde{X}^{s'}_{\ell,j}\widetilde{X}^{s'}_{\ell',i}\right)\\
= \frac{1}{(n_e-1)^2}\sum_{s=1}^{n_e}\sum_{s'=1}^{n_e}\left(\sum_{i=1}^n\widetilde{X}^{s}_{\ell,i}\widetilde{X}^{s'}_{\ell',i}\right)\left(\sum_{i=1}^n\widetilde{X}^{s}_{\ell',i}\widetilde{X}^{s'}_{\ell,i}\right)\\
= \frac{1}{(n_e-1)^2}\sum_{s=1}^{n_e}\left(\sum_{i=1}^n\widetilde{X}^{s}_{\ell,i}\widetilde{X}^{s}_{\ell',i}\right)^2 \\
+ \frac{2}{(n_e-1)^2}\sum_{s=1}^{n_e}\sum_{s'> s}^{n_e}\left(\sum_{i=1}^n\widetilde{X}^{s}_{\ell,i}\widetilde{X}^{s'}_{\ell',i}\right)\left(\sum_{i=1}^n\widetilde{X}^{s}_{\ell',i}\widetilde{X}^{s'}_{\ell,i}\right)
\end{multline}

\paragraph{Fourth term: product of inter-level covariances}
\begin{multline}
\sum_{i=1}^n\sum_{j=1}^n\C\left(X_{\ell,i},X_{\ell',i}\right)\C\left(X_{\ell,j},X_{\ell',j}\right)\approx\\
\sum_{i=1}^n\sum_{j=1}^n \left(\frac{1}{n_e-1}\sum_{s=1}^{n_e}\widetilde{X}^{s}_{\ell,i}\widetilde{X}^{s}_{\ell',i}\right)\left(\frac{1}{n_e-1}\sum_{s'=1}^{n_e}\widetilde{X}^{s'}_{\ell,j}\widetilde{X}^{s'}_{\ell',j}\right)\\
=\left(\frac{1}{n_e-1}\sum_{s=1}^{n_e}\sum_{i=1}^n\widetilde{X}^{s}_{\ell,i}\widetilde{X}^{s}_{\ell',i}\right)^2
\end{multline}

\paragraph{Computing in practice}
In practice, these quantities can be expressed as functions of the $n_e^2L^2$ space averages $\gamma(\ell,s,\ell',s')=\sum_{i=1}^n \widetilde{X}^{s}_{\ell,i}\widetilde{X}^{s'}_{\ell',i}$.
Grouping the four terms together, we retrieve

\begin{multline}
\sum_{i=1}^n\sum_{j=1}^n \mathbb{C}^{(k, ij)}_{\ell,\ell'} = \frac{1}{m^{(k)}}\left(\frac{\sum_{s=1}^{n_e}\gamma(\ell,s,\ell',s)^2}{n_e }
- \frac{\sum_{s=1}^{n_e}\sum_{s'=1}^{n_e}\gamma(\ell,s,\ell',s')^2}{(n_e-1)^2}\right) \\
+ \frac{1}{m^{(k)}{(m^{(k)}-1)}}\left(\frac{\sum_{s=1}^{n_e}\sum_{s'=1}^{n_e}\gamma(\ell,s,\ell',s')\gamma(\ell',s,\ell,s')}{(n_e-1)^2}
+ \frac{\left(\sum_{s=1}^{n_e}\gamma(\ell,s,\ell',s)\right)^2}{(n_e-1)^2}\right)\label{eq:gamma_to_C}
\end{multline}

Note the symmetry $\gamma(\ell,s,\ell',s') = \gamma(\ell',s',\ell,s)$, which reduces to $n_e(n_e+1)L^2/2$ the number of space averages to compute.

A possible algorithm would be:
\begin{enumerate}
\item Generate $n_e$ simulations coupled across all $L$ levels: $\left(\left(X_\ell^s\right)_{\ell=1}^L\right)_{s=1}^{n_e}$.
\item Loop over all ensemble members and all fidelity levels to estimate the $L$ ensemble means $\mu_\ell\in\mathbb{R}^n$. These will be used to estimate fourth-order moments in next step.
\item Double loop over ensemble members and double loop over fidelity levels to estimate the point-wise $\gamma$. Make use of the symmetry property. Space-average.
\item Compute the elements of the averaged covariance matrix \eqref{eq:gamma_to_C}.
\end{enumerate}

\paragraph{Remark}
This is valid in the limit of very large $n_e$.
For finite sizes, these estimates are biased, as mentioned earlier.
Unbiased estimators for these quantities do exist, but are more complex (see for instance \citealp{gerlovina2019moments}).

\clearpage
\section{Optimal sample allocation for an MLMC covariance matrix estimator}
\label{app:opt-alloc-mlmc-cov-mat}
The MLMC estimator for a covariance matrix is a specific (sub-optimal) case of \eqref{eq:sc_covmat}, with weights $\beta^{(1)}=\begin{pmatrix}
1
\end{pmatrix}$ and $\beta^{(k)}=\begin{pmatrix}
1& -1
\end{pmatrix}^\T$ for $2\leq k \leq K$.
The generalized variance of this multilevel estimator is given by \eqref{eq:var_sc_covmat}, which can be estimated in practice following appendix \ref{app:covcov_estimate}.
Minimizing the variance as a function of the sample sizes solves the problem of optimal sample allocation.
Note that the dependence on the sample sizes $\left(m^{(k)}\right)_{k=1}^K$ is hidden in the $\mathbb{C}^{(k, ij)}$.
Note also that the minimization should be done numerically, since the variance involves complex terms such as the inverse of $m^{(k)}\left(m^{(k)} - 1\right)$.
This is quite direct in python, for instance using \texttt{scipy.optimize.minimize} with constraints.

\paragraph{Link with \citet{mycek2019multilevel}}
A simpler but possibly suboptimal sample allocation can be found using the approach proposed by \citet{mycek2019multilevel}.
They do not minimize the exact variance of the estimator, but an upper bound of this variance (their equation 2.31).
Their bound is still written a a sum over the coupling groups, but each term is proportional to $1/m^{(k)}$.
As a consequence, an analytical solution exists for the optimal sample allocation, which makes its determination much faster and simpler\footnote{In the situations we tested, the bounds proposed by \citet{mycek2019multilevel} were quite loose (about a factor 2 above the actual variances). 
This had no impact in the sample allocation though, as their relative evolution among levels was very similar to the true variances, which made them a useful proxy for the actual variance contributions.}.

We extend the bound to the multivariate case, simply by summing over all $i, j$ elements of the covariance matrix to be estimated.
With our notations, the contributions to the variance include the sample size (different from the notations of \cite{mycek2019multilevel}).
\begin{align}
\mathcal{V}_k &= \left(\beta^{(k)}\right)^\T \left(\sum_{i=1}^n\sum_{j=1}^n \mathbb{C}^{(k, ij)}\right)\beta^{(k)}\\
\mathcal{V}_k &\leq \sum_{i=1}^n\sum_{j=1}^n \frac{1}{2}\frac{1}{m^{(k)}-1}\left[\sqrt{\mathbb{M}^4\left[\Delta_{\ell, i}\right]\mathbb{M}^4\left[\Sigma_{\ell, j}\right]}
+ \sqrt{\mathbb{M}^4\left[\Delta_{\ell, j}\right]\mathbb{M}^4\left[\Sigma_{\ell, i}\right]}\right]\\
&= \frac{1}{m^{(k)}-1} \left(\sum_{i=1}^n\sqrt{\mathbb{M}^4\left[\Delta_{\ell, i}\right]}\right)\left(\sum_{i=1}^n\sqrt{\mathbb{M}^4\left[\Sigma_{\ell, i}\right]}\right)
\end{align}
where $\Delta_{\ell,i}=X_{\ell,i} - X_{\ell-1,i}$  and $\Sigma_{\ell,i}=X_{\ell,i} + X_{\ell-1,i}$ (assuming undefined variables are zero).

A possible algorithm would be:
\begin{enumerate}
\item Generate $n_e$ simulations coupled across all $L$ levels.
\item Loop over ensemble members and fidelity levels to estimate the $L$ ensemble means $\mu_\ell\in\mathbb{R}^n$. These will be used to estimate fourth-order moments in next step.
\item Loop over ensemble members and over fidelity levels to estimate the point-wise fourth-order moments $\mathbb{M}^4\left[\Delta_{\ell, i}\right]$ and $\mathbb{M}^4\left[\Sigma_{\ell, i}\right]$.
\item Space-average and multiply to get the variance contribution terms.
\end{enumerate}

\clearpage
\section{Optimal localization using random asymptotic quantities}
\label{app:opt-loc}

This sections details how the results of section \ref{sec:opt-loc} can be derived (more rigorously?) with the formalism of \BMa{} and \BMb.

The main difference consists in considering asymptotic quantities as random, consistently with linear filtering theory.
We assume the existence of two independent random processes $\mathcal{R}_1$ and $\mathcal{R}_2$.
The first process generates asymptotic quantities denoted as $\mathcal{S}$ (for instance first, second and fourth order moments).
The second one generates members consistent with these quantities (typically using an Ensemble of Data Assimilations).

The mean squared error is to be minimized over both process, though we only have access to one realization of $\mathcal{R}_1$.
The expectation in the MSE is thus the expectation over both processes.
We could either index the expectation operators by the processing they refer to, or use the formalism of conditional expectations:

\begin{align}
\mse\left(\widehat{\mathbf{B}}^\text{ML}, \mathbf{B}_L\right) &= \mathbb{E}\left[\left\|\sum_{k=1}^K \sum_{\ell\in S^{(k)}} \mathbf{L}_{\ell}^{(k)} \circ \widetilde{\mathbf{B}}_{\ell}^{(k)} - \mathbf{B}_{L}\right\|^2_\text{F}\right]\\
&= \mathbb{E}_1\left[\E_2\left[\left\|\sum_{k=1}^K \sum_{\ell\in S^{(k)}} \mathbf{L}_{\ell}^{(k)} \circ \widetilde{\mathbf{B}}_{\ell}^{(k)} - \mathbf{B}_{L}\right\|^2_\text{F}\right]\right]\\
&= \mathbb{E}\left[\E\left[\left.\left\|\sum_{k=1}^K \sum_{\ell\in S^{(k)}} \mathbf{L}_{\ell}^{(k)} \circ \widetilde{\mathbf{B}}_{\ell}^{(k)} - \mathbf{B}_{L}\right\|^2_\text{F}\right]\right|\mathcal{S}\right]
\end{align}

The rest of the derivation follows section \ref{sec:optloc_general}.
There is just one additional argument needed to simplify the cross terms $\E\left[B_{L}\underline{\widetilde{{B}}}^\T\right]$ when expanding the MSE:

\paragraph{Expectations of products of sampled and asymptotic quantities}
We have assumed the independence of the sampling error and the process $\mathcal{R}_2$ generating asymptotic statistics.
\begin{align}
\E\left[\left(\widetilde{B}_\ell - B_\ell\right)B_L\right] &= \E\left[\widetilde{B}_\ell - B_\ell\right]\E\left[B_L\right]\\
&=0\\
\mathit{i.e.}\quad \E\left[\widetilde{B}_\ell B_L\right] &= \E\left[B_\ell B_L\right]
\end{align}
So $\E\left[B_{L}\underline{\widetilde{{B}}}^\T\right] = \E\left[B_{L}\underline{B}\right]$.

\paragraph{Interpretation of the ergodicity hypothesis}
The ergodicity assumption intervenes within the same context of equivalence classes.
A sum over a random subset of an equivalence class is meant to approximate the expectations over both random processes $\mathcal{R}_1$ and $\mathcal{R}_2$.

\printbibliography

@Article{schaden2021asymptotic,
  author    = {Daniel Schaden and Elisabeth Ullmann},
  journal   = {{SIAM}/{ASA} Journal on Uncertainty Quantification},
  title     = {Asymptotic Analysis of Multilevel Best Linear Unbiased Estimators},
  year      = {2021},
  month     = jan,
  number    = {3},
  pages     = {953--978},
  volume    = {9},
  doi       = {10.1137/20m1321607},
  publisher = {Society for Industrial {\&} Applied Mathematics ({SIAM})},
}

@Article{schaden2020multilevel,
  author    = {Daniel Schaden and Elisabeth Ullmann},
  journal   = {{SIAM}/{ASA} Journal on Uncertainty Quantification},
  title     = {On Multilevel Best Linear Unbiased Estimators},
  year      = {2020},
  month     = jan,
  number    = {2},
  pages     = {601--635},
  volume    = {8},
  doi       = {10.1137/19m1263534},
  publisher = {Society for Industrial {\&} Applied Mathematics ({SIAM})},
}

@Misc{menetrier2020covariance,
  author    = {M{\'e}n{\'e}trier, Benjamin},
  month     = aug,
  title     = {Sample covariance filtering},
  year      = {2020},
  copyright = {Open Access},
  doi       = {10.5281/ZENODO.4009099},
  publisher = {Zenodo},
}

@Article{menetrier2015linear2,
  author  = {M{\'e}n{\'e}trier, Benjamin and Montmerle, Thibaut and Michel, Yann and Berre, Loïk},
  journal = {Monthly Weather Review},
  title   = {Linear filtering of sample covariances for ensemble-based data assimilation. {Part} {II}: Application to a convective-scale {NWP} model},
  year    = {2015},
  number  = {5},
  pages   = {1644--1664},
  volume  = {143},
  doi     = {10.1175/mwr-d-14-00156.1},
}

@Article{menetrier2015linear1,
  author    = {M{\'e}n{\'e}trier, Benjamin and Montmerle, Thibaut and Michel, Yann and Berre, Loïk},
  journal   = {Monthly Weather Review},
  title     = {Linear Filtering of Sample Covariances for Ensemble-Based Data Assimilation. Part {I}: Optimality Criteria and Application to Variance Filtering and Covariance Localization},
  year      = {2015},
  month     = may,
  number    = {5},
  pages     = {1622--1643},
  volume    = {143},
  doi       = {10.1175/mwr-d-14-00157.1},
  publisher = {American Meteorological Society},
  url       = {https://doi.org/10.1175/mwr-d-14-00157.1},
}

@Article{mycek2019multilevel,
  author     = {Mycek, Paul and De Lozzo, Matthias},
  journal    = {{SIAM}/{ASA} Journal on Uncertainty Quantification},
  title      = {Multilevel {M}onte {C}arlo Covariance Estimation for the Computation of {S}obol' Indices},
  year       = {2019},
  month      = jan,
  number     = {4},
  pages      = {1323--1348},
  volume     = {7},
  doi        = {10.1137/18m1216389},
  publisher  = {Society for Industrial {\&} Applied Mathematics ({SIAM})},
  readstatus = {read},
}

@Article{giles2008multilevel,
  author    = {Michael B. Giles},
  journal   = {Operations Research},
  title     = {Multilevel {M}onte {C}arlo Path Simulation},
  year      = {2008},
  month     = jun,
  number    = {3},
  pages     = {607--617},
  volume    = {56},
  doi       = {10.1287/opre.1070.0496},
  publisher = {Institute for Operations Research and the Management Sciences ({INFORMS})},
}

@Article{giles2015multilevel,
  author     = {Michael B. Giles},
  journal    = {Acta Numerica},
  title      = {Multilevel {M}onte {C}arlo methods},
  year       = {2015},
  month      = apr,
  pages      = {259--328},
  volume     = {24},
  doi        = {10.1017/s096249291500001x},
  publisher  = {Cambridge University Press ({CUP})},
  readstatus = {read},
}

@Article{peherstorfer2018survey,
  author    = {Peherstorfer, Benjamin and Willcox, Karen and Gunzburger, Max},
  journal   = {SIAM Rev.},
  title     = {Survey of Multifidelity Methods in Uncertainty Propagation, Inference, and Optimization},
  year      = {2018},
  issn      = {0036-1445},
  month     = jan,
  number    = {3},
  pages     = {550--591},
  volume    = {60},
  comment   = {doi: 10.1137/16M1082469},
  doi       = {10.1137/16M1082469},
  publisher = {Society for Industrial and Applied Mathematics},
  url       = {https://doi.org/10.1137/16M1082469},
}

@Article{gorodetsky2020generalized,
  author   = {Gorodetsky, Alex A. and Geraci, Gianluca and Eldred, Michael S. and Jakeman, John D.},
  journal  = {Journal of Computational Physics},
  title    = {A generalized approximate control variate framework for multifidelity uncertainty quantification},
  year     = {2020},
  issn     = {0021-9991},
  pages    = {109257},
  volume   = {408},
  doi      = {10.1016/j.jcp.2020.109257},
  url      = {https://www.sciencedirect.com/science/article/pii/S0021999120300310},
}

@Article{lorenc2003potential,
  author    = {Lorenc, Andrew C.},
  journal   = {Quarterly Journal of the Royal Meteorological Society},
  title     = {The potential of the ensemble {Kalman} filter for {NWP} — a comparison with {4D}-{V}ar},
  year      = {2003},
  number    = {595},
  pages     = {3183--3203},
  volume    = {129},
  doi       = {10.1256/qj.02.132},
  publisher = {Wiley Online Library},
}

@Article{buehner2005ensemble,
  author     = {Buehner, Mark},
  journal    = {Quarterly Journal of the Royal Meteorological Society},
  title      = {Ensemble-derived stationary and flow-dependent background-error covariances: {Evaluation} in a quasi-operational {NWP} setting},
  year       = {2005},
  number     = {607},
  pages      = {1013--1043},
  volume     = {131},
  doi        = {10.1256/qj.04.15},
  printed    = {printed},
  publisher  = {Wiley Online Library},
  ranking    = {rank5},
  readstatus = {skimmed},
}

@Article{bannister2017review,
  author    = {Bannister, Ross Noël},
  journal   = {Quarterly Journal of the Royal Meteorological Society},
  title     = {A review of operational methods of variational and ensemble-variational data assimilation},
  year      = {2017},
  number    = {703},
  pages     = {607--633},
  volume    = {143},
  doi       = {10.1002/qj.2982},
  printed   = {printed},
  publisher = {Wiley Online Library},
}

@Article{menetrier2015optimized,
   author = {Benjamin Ménétrier and Thomas Auligné},
   doi = {10.1175/MWR-D-15-0057.1},
   issn = {0027-0644},
   issue = {10},
   journal = {Monthly Weather Review},
   month = {10},
   pages = {3931-3947},
   title = {Optimized Localization and Hybridization to Filter Ensemble-Based Covariances},
   volume = {143},
   url = {http://journals.ametsoc.org/doi/10.1175/MWR-D-15-0057.1},
   year = {2015},
}

@Article{gerlovina2019moments,
author = {Inna Gerlovina and Alan E. Hubbard},
editor = {Yan Sun},
title = {Computer algebra and algorithms for unbiased moment estimation of arbitrary order},
journal = {Cogent Mathematics \& Statistics},
volume = {6},
number = {1},
pages = {1701917},
year  = {2019},
publisher = {Cogent OA},
doi = {10.1080/25742558.2019.1701917},
URL = {https://doi.org/10.1080/25742558.2019.1701917},
}

@article{croci2023multifidelity,
archivePrefix = {arXiv},
arxivId = {2301.07831},
author = {Croci, M. and Willcox, K. E. and Wright, S. J.},
eprint = {2301.07831},
month = jan,
title = {{Multi-output multilevel best linear unbiased estimators via semidefinite programming}},
url = {http://arxiv.org/abs/2301.07831},
year = {2023}
}

\end{document}